\documentclass[a4paper,11pt]{article}
\usepackage[utf8]{inputenc}
\usepackage[english]{babel}
\usepackage{amsmath, amssymb,graphicx}
\pdfoutput=1 
\usepackage{jheppub} 
\newcommand{\be}{\begin{equation}}
\newcommand{\ee}{\end{equation}}
\newcommand{\bea}{\begin{eqnarray}}
\newcommand{\eea}{\end{eqnarray}}
\newcommand{\nn}{\nonumber}

\title{\boldmath Shock waves in Lifshitz-like spacetimes}

\author[a]{Irina Ya. Aref'eva}
\author[b]{and Anastasia A. Golubtsova}

\affiliation[a]{Steklov Mathematical Institute, Russian Academy of Sciences,\\Gubkina str. 8, 119991, Moscow, Russia}
\affiliation[b]{Bogoliubov Laboratory of Theoretical Physics, JINR,\\141980  Dubna, Moscow region, Russia}

\emailAdd{arefeva@mi.ras.ru}
\emailAdd{golubtsova@theor.jinr.ru}

\abstract{We construct shock waves for Lifshitz-like geometries in four- and
five-dimensional effective theories as well as in D3-D7 and D4-D6 brane systems. The
solutions to the domain wall profile equations are found. Further, the study makes  a connection with the implications 
for the quark-gluon plasma formation in heavy-ion collisions.  According to the holographic approach, 
the multiplicity of particles produced in heavy-ion collisions can be estimated by the
area of the trapped surface formed in shock wave collisions. 
We calculate the areas of trapped surfaces  in the geometry of two colliding Lifshitz domain walls. Our estimates show that for five-dimensional cases with
 certain values of the critical exponent the dependence of multiplicity 
on the energy of colliding ions is rather close to the experimental data ${\cal M} \sim s^{\,0.15}$ observed at RHIC and LHC. }


\begin{document}
\maketitle
\flushbottom
\newpage
\section{Introduction}
\label{sec:intro}

 In recent years, the original AdS/CFT correspondence \cite{Malda,GKP,Witten} has gone through a number of transformations (for reviews of phenomenological applications see \cite{Solana,IA,DeWolf}). Its extensions
have been explored to yield insights into strongly coupled field theories different from a textbook example of $\mathcal{N} = 4$ super Yang-Mills theory. 
In this context, a holographic duality between gravitation and field theories with a Lifshitz scaling symmetry  has recently received much attention.
According to the conjectured duality \cite{KLM} the symmetries of the gravitational background
\begin{equation}\label{1.1}
ds^{2} = L^{2}\left[ - r^{2\nu}dt^{2} + r^{2}d\vec{x}^{2}_{d-1} + \frac{dr^{2}}{r^{2}} \right],
\end{equation}
with the so-called dynamical Lifshitz exponent $\nu$\footnote{The standard notation of the dynamical exponent is $z$. Here we define the critical exponent by $\nu$ to avoid the confusion with the holographic coordinate later.} realize the symmetries of the dual field theory invariant under the scale transformations
\begin{equation}\label{1.2}
t \rightarrow \lambda^{\nu}t, \quad \vec{x} \rightarrow \lambda \vec{x}, \quad r \rightarrow \frac{1}{\lambda}r.
\end{equation}

The concept of the anisotropic scaling \eqref{1.2} came from the condensed matter physics and characterizes  the so-called Lifshitz fixed points \cite{RMH}. These critical points arise in the phase diagrams of various physical systems  \cite{HWD}. The appearance of  Lifshitz points can be understood within the Landau theory theory of phase transitions \cite{Yuk,LL}.

The dynamical exponent $\nu =1$ describes the metric of the anti-de Sitter spacetime  and the scaling symmetry \eqref{1.2} turns to be the usual one embedded in the conformal group $SO(d+1,2)$, so the scaling is isotropic and corresponds to relativistic invariance. 
The metric (\ref{1.1}) is not a solution to the vacuum Einstein equations --
 the simplest matter content required  to support Lifshitz geometry includes a negative cosmological constant and massive vector fields  \cite{KLM}.
It is worth noting that the case $\nu < 1$ produces an unrealistic causal structure in the field theory and corresponds to violating the null energy condition \cite{HK}.

All curvature invariants built from the Riemann tensor for \eqref{1.1} are finite constants. 
However,  for all $\nu > 1$ the Lifshitz metrics are singular as  $r \rightarrow 0$  in the sense of pp-curvature singularities, which can be found out by computing the tidal forces between infalling geodesics.
For \eqref{1.1} it was pointed in \cite{KLM} and demonstrated explicitly in \cite{CM, HW}. The possible way of the resolution of singularities  has been outlined in \cite{BDHS, HKW}.

Embedding Lifshitz spacetimes into string theory and supergravity seems to be a non-trivial problem.
However, explicit solutions with Lifshitz symmetries with certain dynamical exponents  were presented in works \cite{BN}-\cite{GPTZ}. 
Black branes/holes in asymptotically Lifshitz spacetimes were constructed in \cite{MT1}-\cite{TV}.
The global structure of Lifshitz black holes with arbitrary exponent $\nu >1$ was analyzed in \cite{BDTZ}.

Since the anisotropic scaling (\ref{1.2}) takes place in a number of physical finite density systems, Lifshitz spacetimes found  its natural applications 
to holographic description of several condensed matter phenomena, namely superfluidity and superconductivity (see \cite{SAH}-\cite{CGKKM} and refs. therein). 
The dual description of condensed matter systems at finite temperatures is represented  by Lifshitz black hole (brane) solutions.

There exists a number of generalizations of the Lifshitz spacetime (\ref{1.1}). Recently, the hyperscaling violating Lifshitz metric has been 
proposed in \cite{CGKKM}-\cite{DHKYW}
\be\label{1.2a}
ds^{2} = r^{-\frac{2\theta}{d}}\left(-r^{2z}dt^{2} + r^{2}d\vec{x}^{2}_{d-1} + \frac{dr^{2}}{r^{2}} \right),
\ee
 which is characterized by the hyperscaling violation parameter $\theta$ in addition to the dynamical exponent $\nu$.

In \cite{SSP} Lifshitz geometries (\ref{1.1}) were generalized to anisotropic Lifshitz-like ones  with the metric
\begin{equation}\label{1.3}
ds^{2} = L^{2}\left(r^{2\nu}\left(-dt^{2} + \sum^{p}_{i = 1}dx^{2}_{i}\right) + r^{2} \sum^{q}_{j =1}dy^{2}_{j} + \frac{dr^{2}}{r^{2}}\right),
\end{equation}
where $p > 0$, $q > 0$ and $p + q+2 =D \equiv d+ 2$. We will refer to these metrics as Lifshitz-like metrics {\bf(p,q)}-type ($Lif_{(p,q)}$). The line element (\ref{1.3}) is invariant under a generalized scaling
\begin{equation}\label{1.4}
(t, x_{i}, y_{i}, r) \rightarrow \left(\lambda^{\nu}t, \lambda^{\nu}x_{i}, \lambda y_{i}, \frac{r}{\lambda}\right).
\end{equation}
 The scaling (\ref{1.4}) differs from (\ref{1.2}) by the extension of the anisotropy for space coordinates $x_{i}$ and 
 thus, the symmetry of the spatial part is broken down to $SO(p) \times SO(q)$. 
 Lifshitz-like fixed points with the anisotropic scaling (\ref{1.4}) can arise in  magnetic systems \cite{HWD}.

String embedding of the Lifshitz-like spacetime (\ref{1.3}) with the anisotropic scaling (\ref{1.4}) and the dynamical exponent $\nu = 3/2$  was developed for  the system of intersecting $D3-D7$ branes in \cite{ALT}, where its finite temperature extension has also been presented. Holographic calculations of the thermal and the entanglement entropies corresponding to these solutions were demonstrated that both quantities enjoy characteristic scaling properties \cite{ALT}.

In \cite{MT,MT2} the finite-temperature generalization of the type IIB supergravity solution of \cite{ALT} was studied as a gravity 
dual to an anisotropic deformation of a four-dimensional $\mathcal{N} = 4$ super Yang-Mills theory. 
It was noted in \cite{MT2} that at zero temperature  the metric has a naked curvature singularity deep in the IR.
The existence of solutions which interpolate between the anisotropic solutions in the IR and the  $AdS_5 \times X_5$ solutions in the UV was shown in \cite{ALT, MT2}.
 These interpolating solutions can be considered as the dual of the RG flow between the two systems \cite{MT,MT2}.

In this paper, we consider Lifshitz-like backgrounds in the context of their  applications to the anisotropic quark-gluon plasma created in heavy-ion collisions \cite{MT}-\cite{RS}.
A holographic model with a Lifshitz-like spacetime in the IR and AdS boundary conditions is supposed to be related with an anisotropic SYM quark-gluon plasma \cite{MT2}. 
 Several physical quantities have been explored in this frameworks.
Namely, the shear viscosity to the entropy ratio  $\eta/s$ for this model was explored in \cite{RS2}, where a violation of the usual holographic bound of gravity duals by certain components of the viscosity tensor was found. Anisotropy effects on heavy-quark energy loss in an anisotropic plasma were studied in \cite{Fada},  
the energy loss due to radiation in the anisotropic case has been shown to be  less than the isotropic one.
Studies of  jet quenching,  drag force and static potential  for a strongly coupled anisotropic plasma, described by the Lifshitz-like background from 
 \cite{MT2},  include \cite{CFMT}-\cite{CFMT2}. It has been shown \cite{CFMT, CFMT2} that the drag coefficient and the jet quenching parameter 
can be larger or smaller than its corresponding isotropic value depending on the initial conditions.

One should mention another attempt to describe an anisotropic plasma presented in \cite{JW}. This approach is based  on 
the boundary field theory with anisotropic pressures and its gravity dual involving a benign naked singularity. Opposed to the model \cite{MT} it does not have a hydrodynamical limit. In \cite{RS} the effects of anisotropy on  heavy quark potentials and jet quenching parameters were compared for both holographic models. According to the comparison the holographic models have a quite different behaviour at large distances in the anisotropic direction.

Owing to the holographic approach, the creation and evolution processes of the quark-gluon plasma can be examined through the analysis both shock waves \cite{GPY}-\cite{APP} and Vaidya solutions \cite{BBBC}-\cite{AV}. It should be noted that the Lifshitz-Vaidya solution has been constructed in \cite{KKVT}. This spacetime describes for a shell falling at the speed of light and provides a holographic model of a quench near a quantum critical point. 

Here we aim to construct shock waves in the Lifshitz-like spacetime (\ref{1.3}) with an arbitrary dynamical exponent $\nu$. 
There are two approaches to construct shock waves solutions in (A)dS spacetime. One of these consists in considering  a boosted black hole \cite{HT}, 
the second deals with a direct method for constructing solutions to E.O.M. with a  
source containing  a $\delta$-function located at zero of a light-cone coordinate.  
For the case of (A)dS both methods give the same results.  
The appearance of $\delta$-functions as a source can be treated in the distribution language \cite{ABJ-gf}. 

In this paper we show that the shock waves in the $Lif_{(p,q)}$-background (\ref{1.3}) satisfy the following equation\footnote{For the D-dimensional Lifshitz-like metric  we give a derivation of eqs. (\ref{1.5}), (\ref{1.7}) and the solution for the profile in Appendix~\ref{App:A}.}
\begin{equation}\label{1.5}
\left[\Box_{Lif_{(p+q)}} - \frac{1}{L^{2}}\left(p + \frac{q}{\nu}\right)\right]\frac{\phi(x_{i},y_{j},z)}{z} = -2z  J_{uu},
\end{equation}
where $i = 1, \ldots p-1, j  =1, \ldots q$, $u$ is one of the light-cone coordinates, $J_{uu}$  is the  density related with the stress-energy tensor, 
$T_{uu}\sim J_{uu}\delta(u)$
 ($T_{uu}$  is the only non-zero component of the stress-energy tensor)  and $\Box_{Lif_{(p+q)}}$ is the Laplace-Beltrami operator defined on the $(p+q)$-dimensional Lifshitz-like space with the metric
\begin{equation}\label{1.6}
ds^{2}_{Lif_{(p+q)}} = \frac{1}{z^{2}}\sum^{p - 1}_{i =1}dx^{2}_{i} + \frac{1}{z^{2/\nu}}\sum^{q}_{j =1}dy^{2}_{j} + \frac{dz^{2}}{z^{2}},
\end{equation}
with the redefined coordinate $z = r^{-\nu}$.

To simplify the holographic description of heavy-ion collisions one can use shock domain walls \cite{LS,ABP,APP}. 
In this case eq. (\ref{1.5}) is reduced to the following form
\begin{equation}\label{1.7}
\frac{\partial^{2}\phi(z)}{\partial z^{2}} - \left(p + \frac{q}{\nu}\right)\frac{1}{z}\frac{\partial \phi(z)}{\partial z} = -16 \pi G_{D}Ez^{p + \frac{q}{\nu}}_{*}\delta(z - z_{*}).
\end{equation}

Here we  study wall-on-wall collisions in  backgrounds with the Lifshitz-like scaling.  As it is known, the quark-gluon plasma is anisotropic at the very early stages of heavy-ion collisions, in which most of the entropy is produced. In holography, the entropy production is related to the trapped surface formed during the shock waves (domain walls) 
collisions.  Following this proposal, we will calculate the area of the trapped surface for the wall-on-wall collision and estimate the multiplicity of particle production.

The paper is organized as follows. 

In Sect.~\ref{Sec:2} we present the construction of shock waves in 4d Lifshitz-like spacetimes.
We show that the shock wave in this case is located on a 2d Lifshitz space and the shock wave profile up to a rescaling 
factor is a fundamental solution  to the Laplace-Beltrami equation with a non-zero cosmological constant in the 2d Lifshitz space, eq.(\ref{3.9}).  
In Sect.~\ref{Sec:2.3} the shape for the shock domain wall profile is obtained. In Sect.~\ref{Sect:2.4}  the collision of two domain walls
is considered.  We find the conditions for the trapped surface formation and estimate the multiplicity. 

In Sect.~\ref{Sect:3} we develop the similar technic in 5d Lifshitz-like spacetimes. In Sect.~\ref{Sect:3.1} we construct the shock wave geometries in type {\bf(1,2)} 5-dimensional Lifshitz-like backgrounds and the same for type {\bf(2,1)} 5-dimensional Lifshitz-like backgrounds in Sect.~\ref{Sect:3.2}.  

In Sect.~\ref{Sect:4} we study shock waves, domain walls and their collisions for D3-D7 and D4-D6 brane systems 
applying the approach employed in previous sections. 

We conclude in Sect.~\ref{Sect:5} with a discussion of our results. 
In appendices we collect some technical details of computations used for constructing shock waves.

\section{Shock waves in $3+1$-dimensional Lifshitz-like spacetimes} \label{Sec:2}

In this section, we will consider the construction of shock wave solutions in  a four-dimensional Lifshitz-like spacetime. 
Our starting point  is the 4-dimensional model proposed in \cite{SSP}. Here we have $p=1$, $q=1$ for the four-dimensional metric  (\ref{1.3}).

\subsection{$3+1$-dimensional model and Lifshitz-like solutions}

First we provide a brief overview of characteristics corresponding to Lifshitz-like solutions (\ref{1.3}).
We consider a 4-dimensional effective gravity model governed by the action \cite{SSP}
\begin{eqnarray}\label{2.1a}
S = \frac{1}{2\kappa^{2}}\int d^{4}x\Bigl[ \sqrt{|g|}\left(R - 2\Lambda  - \frac{F^{2}_{(2)}}{4}  - \frac{F^{2}_{(3)}}{12} - \frac{H^{2}_{(3)}}{12} - \frac{m^{2}_{0}}{2} B^{2}_{(2)}\right) - \nonumber\\ c\epsilon^{M_{1}M_{2}M_{3}M_{4}}A_{M_{1}M_{2}}F_{M_{3}M_{4}}\Bigr],
\end{eqnarray}
where $B_{(2)}, F_{(2)}, F_{(3)}, H_{(3)}$ are form fields , $F_{(2)} = dA_{(1)}$, $H_{(3)}=dB_{(2)}$, 
$c$ is the topological coupling between the two and three form fluxes  and $\Lambda$ is the negative cosmological constant.

The Einstein equations  for (\ref{2.1a}) are given by
\begin{eqnarray}\label{2.1b}
R_{MN} - \frac{1}{2}g_{MN}R + g_{MN}\Lambda = \frac{1}{2}F_{MM_{1}}F^{M_{1}}_{N} + \frac{1}{4}H_{MM_{1}M_{2}}H^{M_{1}M_{2}}_{N} + m^{2}_{0}B_{MM1}B_{N}^{M_{1}} + \nonumber\\ 
\frac{1}{4}F_{MM_{1}M_{2}}F_{N}^{M_{1}M_{2}} -  \frac{1}{2}g_{MN}\left[\frac{F^{2}_{(2)}}{4} + \frac{F^{2}_{(3)}}{12} + \frac{H^{2}_{(3)}}{12} + \frac{m^{2}_{0}}{2}B^{2}_{(2)}\right].
\end{eqnarray}

The  equations of motions for the field strengths read
\begin{eqnarray}\label{2.1c}
\partial_{M_{3}}\left(\sqrt{|g|}F^{M_{3}M_{4}}\right) = - \frac{2}{3}\epsilon^{M_{1}M_{2}M_{3}M_{4}}cF_{M_{1}M_{2}M_{3}}, \\
\partial_{M_{3}}\left(\sqrt{|g|}F^{M_{1}M_{2}M_{3}}\right) =  2 c\epsilon^{M_{1}M_{2}M_{3}M_{4}}F_{M_{3}M_{4}}, \\ \label{2.1d}
\partial_{M_{3}}\left(\sqrt{|g|}H^{M_{1}M_{2}M_{3}}\right) =  2 m^{2}_{0}\sqrt{|g|}B^{M_{1}M_{2}}.
\end{eqnarray}

Eqs. (\ref{2.1c})-(\ref{2.1d}) allow  the following anisotropic metric ansatz of our interest 
\begin{equation}\label{2.1}
ds^{2} = L^{2}\left(r^{2\nu}\left(-dt^{2} +  dx^{2}\right) + r^{2} dy^{2} + \frac{dr^{2}}{r^{2}}\right),
\end{equation}
where the critical exponent $\nu$ is arbitrary.  The geometry (\ref{2.1}) is sourced by two and three form-fluxes $B_{(2)}$ and $H_{(3)}$ given by
\begin{equation}\label{2.0a}
B_{(2)} = \sqrt{\frac{\nu -1}{\nu}}L^{2}r^{2\nu} dt \wedge dx, \quad H_{(3)} = 2\nu\sqrt{\frac{\nu-1}{\nu}}L^{2}r^{2\nu - 1} dr\wedge dt\wedge dx
\end{equation}
and the constants obey  the following constraints
\begin{equation}\label{2.0b}
m^{2}_{0} = \frac{\nu}{L^{2}} , \quad c^{2} = \frac{(\nu +1)\nu}{16L^{2}} , \quad \Lambda = - \frac{4\nu^{2} + \nu +  1}{2L^{2}}.
\end{equation}
Introducing the coordinate redefinition $r^{\nu} = \rho$, one can rewrite the metric (\ref{2.1}) in the following form \cite{ALT}

\begin{equation}\label{2.2}
ds^{2} = L^{2} \left[\rho^{2}\left(-dt^{2} + dx^{2}\right) + \rho^{2/\nu}dy^{2} +  \frac{d\rho^{2}}{\rho^{2}}\right].
\end{equation}

The background (\ref{2.2}) is invariant under the  transformation
\begin{equation}\label{2.2b}
(t, x, y, \rho) \rightarrow \left(\lambda t, \lambda x, \lambda^{1/\nu}y, \frac{\rho}{\lambda}\right).
\end{equation}
Thus, the $y$ direction is responsible for the Lorentz symmetry violation and anisotropy.

The Lifshitz geometry (\ref{2.2}) exhibits spacetime isometries. These isometries are generated by the following Killing vectors
\begin{eqnarray}\label{2.4}
\xi = -\frac{\partial}{\partial t}, \zeta_{1} =  \frac{\partial}{\partial x}, \zeta_{2} = \frac{\partial}{\partial y}, 
\eta = -t\frac{\partial}{\partial x} - x \frac{\partial}{\partial t}, 
 \chi = - x\frac{\partial}{\partial x} - t\frac{\partial}{\partial t} - \frac{y}{\nu}\frac{\partial}{\partial y} + \rho\frac{\partial}{d \rho}.
\end{eqnarray}

In terms of the  variable  $z = 1/\rho$ the metric (\ref{2.2}) can be rewritten as follows
\begin{equation}\label{2.2a}
ds^{2} =L^{2}\left[\frac{\left(-dt^{2} + dx^{2}\right)}{z^{2}} + \frac{dy^{2}}{z^{2/\nu}} +  \frac{d z^{2}}{z^{2}}\right].
\end{equation}
It is worth noting that for $\nu = 1$ the spacetime (\ref{2.2a}) reduces to the Poincare patch of $AdS_{4}$.

\subsection{ Shock waves in $3+1$-Lifshitz-like spacetimes} 

To obtain shock waves here and what follows we use the approach from the work  \cite{KS}.  There it is shown that for the $d$-dimensional metric
\be\label{3.1b}
ds^{2} = A(u,v) dudv + g(u, v)h_{ij}(x)dx^{i}dx^{j},
\ee
with $i,j, = 1,2, \ldots, d-2,$ sourced some matter fields and the cosmological constant, one can represent a shock wave solution 
as a metric of a light-like particle located at $u = 0$ and moving with the speed of light in the $v$-direction in the background (\ref{3.1b})
\be\label{3.1c}
ds^{2} = 2 Adudv - 2 Af\delta (u)du^{2} + g h_{ij}dx^{i}dx^{j},
\ee
with the only non-zero component $T_{uu}$ of the stress-energy tensor
\be\label{3.1d}
T_{uu} = -4 p A^{2}\delta^{d-2}(x)\delta (u),  
\ee
where $\delta$ is the Dirac delta-function and $p$ is the momentum of the particle.

Thus, to construct the shock wave in the background (\ref{2.2a}) it is convenient to introduce the null combinations 
\begin{equation}\label{3.2}
du = dt - dx, \quad dv = dt + dx.
\end{equation}
In terms of  $uv$-coordinates  the metric (\ref{2.2a}) can be rewritten in the following form
\begin{equation}\label{3.3}
ds^{2} =L^{2} \left[- \frac{dudv}{z^{2}} + \frac{dy^{2}}{z^{2/\nu}} +  \frac{dz^{2}}{z^{2}}\right].
\end{equation}
As it was pointed earlier, the shock wave solution can be obtained considering the metric (\ref{3.3}) in presence of a light-like particle
\begin{equation}\label{3.4a}
ds^{2} = L^{2}\left[\frac{\phi(y,z)\delta(u) }{z^{2}}du^{2} - \frac{dudv}{z^{2}} + \frac{dy^{2}}{z^{2/\nu}} +  \frac{dz^{2}}{z^{2}}\right].
\end{equation}
Non-zero components of the Ricci tensor and the scalar curvature corresponding to the metric (\ref{3.4a}) read
\begin{eqnarray}\label{3.6c}
R_{uu} = - \frac{1}{2}\frac{\delta(u)}{z\nu}\Bigl[z\nu\frac{\partial^{2} \phi(y,z)}{\partial z^{2}} +  z^{-1+ \frac{2}{\nu}}\nu\frac{\partial^{2}\phi(y,z)}{\partial y^{2}}  
- \frac{\partial \phi(y,z)}{\partial z}\nu  -  \frac{\partial \phi(y, z)}{\partial z} + \nonumber \\ \label{3.6d}
\frac{4\phi(y,z) \nu}{z} + \frac{2\phi(y,z)}{z}  \Bigr],
\end{eqnarray}
\begin{eqnarray}\label{3.7}
R_{uv} = \frac{2\nu+ 1}{2\nu z^{2}}, \quad  R_{yy} = - \frac{z^{-\frac{2}{\nu}}(2\nu + 1)}{\nu^{2}}, \quad R_{zz} = - \frac{2\nu^{2} + 1}{\nu^{2}z^{2}}, \quad \label{3.7a}
R = - \frac{2(3\nu^{2} + 2\nu + 1)}{L^{2} \nu^{2}}. \nn \\
\end{eqnarray}
To obtain the equation for the shock wave profile, let us consider the $uv$-component of the Einstein equations
\begin{equation}\label{Eina}
R_{uv} - \frac{1}{2}g_{uv}R + g_{uv}\Lambda = T_{uv}.
\end{equation}
Substituting $R_{uv}$ from  (\ref{3.7}), $g_{uv}$ from (\ref{3.4a}) and taking into account that the only non-zero component of the stress-energy tensor is $T_{uu}$, one obtains
\begin{equation}\label{Einb}
\frac{2\nu+1}{2\nu z^{2}} -  \frac{1}{4z^{2}}\frac{2L^{2}(3\nu^{2} + 2\nu + 1)}{L^{2}\nu^{2}} - \frac{L^{2}\Lambda}{2z^{2}} = 0,
\end{equation}
which yields the relation for the cosmological constant $\Lambda$
\begin{equation}\label{Lambda}
\Lambda = -\frac{1}{L^{2}}\left(1 + \frac{1}{\nu} + \frac{1}{\nu^{2}}\right).
\end{equation}

Owing to relations for $R_{uu}$  (\ref{3.6c}) and $g_{uu}$ from (\ref{3.4a}) one can derive the equation for the shock wave profile can be obtained from the $uu$-component of Einstein equations
\begin{equation}\label{3.9}
\left[\Box_{Lif_{2}} - \frac{1}{L^{2}}\left(1 + \frac{1}{\nu}\right)\right]\frac{\phi(y,z)}{z} = -2z t_{uu},
\end{equation}
where $t_{uu}$ is related with $T_{uu}$ via $T_{uu} = t_{uu}\delta(u)$ and the operator $\Box_{Lif_{2}}$ has the following form
\begin{equation}\label{3.6a}
\Box_{Lif_{2}} =  \frac{1}{\nu L^{2}}\left(z^{2}\nu\frac{\partial^{2}}{\partial z^{2}} + \nu z\frac{\partial}{\partial z} - z\frac{\partial}{\partial z} + z^{2/\nu}\nu \frac{\partial^{2}}{\partial y^{2}} \right) 
\end{equation}
and is defined on the space
\begin{equation}\label{3.5b}
ds^{2} = L^{2}\left[\frac{dy^{2}}{z^{2/\nu}} +  \frac{dz^{2}}{z^{2}}\right]
\end{equation}
with Killing vectors given by
\begin{eqnarray}\label{3.5c}
 \zeta =  \frac{\partial}{\partial y}, \quad  \eta = \frac{y}{\nu}\frac{\partial}{\partial y} + z\frac{\partial}{\partial z}, \quad \xi = -\frac{z^{2/\nu}\nu^{2} - y^{2} }{2\nu}\frac{\partial}{\partial y}  + y z \frac{\partial}{\partial z}.
 \end{eqnarray}
It acts on the profile function $\displaystyle{\frac{\phi(y,z)}{z}}$ as
\begin{eqnarray}\label{3.6b}
\Box_{Lif_{2}}\left[\frac{\phi(y,z)}{z}\right] = \frac{1}{\nu L^{2}} \Bigl[z\nu \frac{\partial^{2} \phi (y,z)}{\partial z^{2}} + z^{-1+ \frac{2}{\nu}}\nu\frac{\partial^{2}\phi(y,z)}{\partial y^{2}}   -  \nu \frac{\partial \phi (y,z)}{\partial z} -  \nonumber \\ \frac{\partial \phi(y, z)}{\partial z} +  \frac{\nu}{z} \phi(y,z) + \frac{1}{z}\phi(y,z) \Bigr].
\end{eqnarray}
The shape of the shock wave $\Phi = \displaystyle{\frac{\phi}{z}}$  is given by the solution to (\ref{3.9}) with
\be\label{3.5d}
t_{uu} = 8\pi G_{4} E \left(\frac{z}{L}\right)^{1 + \frac{1}{\nu}}\delta(z - z_{*})\delta(y).
\ee

It is easy to see that for the case $\nu =1$, which corresponds to the four-dimensional $AdS$ space, eq. (\ref{3.9}) comes to the well-known equation for the $4d$ $AdS$-shock wave \cite{AGSTV, AB4}

\begin{equation}\label{AdS4}
\left[\Box_{H_{2}} - \frac{2}{L^{2}}\right]\frac{\phi(y,z)}{z} = -2z t_{uu}.
\end{equation}

\subsection{Shock domain-wall} \label{Sec:2.3}
To find solutions to (\ref{3.9}) with (\ref{3.5d}) looks  to be rather complicated.
In works \cite{LS,ABP}  a simpler form of shock waves called domain walls was suggested. 
To derive the  equation for the profile, one should consider the mass of a point-like source averaged over the domain-wall.
The profile of the domain wall has the dependence only on the holographic coordinate $z$ and obeys the equation
\begin{equation}\label{3.1.1}
\left[\Box_{Lif_{2}} - \frac{1}{L^{2}}\left(1 + \frac{1}{\nu}\right)\right]\frac{\phi(z)}{z} = -16\pi G_{4}z J_{uu}.
\end{equation}

Eq. (\ref{3.1.1}) is reduced to the following form
\begin{equation}\label{3.1.1b}
\frac{\partial^{2} \phi(z)}{\partial z^{2}}  -  \left(1 + \frac{1}{\nu}\right)\frac{1}{z} \frac{\partial \phi(z)}{\partial z} = -16\pi G_{4} J_{uu},
\end{equation}
where  the source is chosen as
\begin{equation}\label{3.1.1c}
J_{uu} = E \left(\frac{z}{L}\right)^{1+1/\nu}\delta(z  - z_{*}).
\end{equation}

The solution to eq. (\ref{3.1.1b}) for the domain wall profile  can be written down in the following form
\begin{equation}\label{3.2.1a}
\phi =\phi_{a}\Theta(z_{*} - z) +\phi_{b}\Theta(z - z_{*}),
\end{equation}
where $z_{*}$ is a constant,  $\Theta(z - z_{*})$ is the Heaviside function and the profile functions are 
\bea\label{3.2.2}
\phi_{a}(z) &=& C_{0}z_{a}z_{b}\left(\frac{z^{(2\nu+1)/\nu}_{*}}{z^{(2\nu+1)/\nu}_{b}} - 1\right)\left(\frac{z^{(2\nu+1)/\nu}}{z^{(2\nu+1)/\nu}_{a}} - 1\right),\\
\phi_{b}(z)& = &C_{0}z_{a}z_{b}\left(\frac{z^{(2\nu+1)/\nu}_{*}}{z^{(2\nu+1)/\nu}_{a}} - 1\right)\left(\frac{z^{(2\nu+1)/\nu}}{z^{(2\nu+1)/\nu}_{b}} - 1\right),\\
C_{0}& =&  -\frac{16 \nu\pi G_{4} E z^{1+1/\nu}_{a}z^{1+1/\nu}_{b}}{(2\nu+1)L^{\frac{1}{\nu} + 3}(z^{(2\nu+1)/\nu}_{b} - z^{(2\nu+1)/\nu}_{a})}.
\eea
It is worth noting that the solution (\ref{3.2.1a}) decreases in both directions from the point $z_{*}$. The point $z_{*}$ can be considered as the center of the shock domain wall.

\subsection{Wall-on-wall collision}\label{Sect:2.4}

Let us discuss the collision of two shock domain walls, as a model of heavy ion collisions. 
According to the proposal of \cite{GPY}, a collision of two nuclei in the bulk can be interpreted as 
a line element for two identical shock waves propagating towards one another in the gravity dual background.  
Here following \cite{LS,ABP} we consider the collision of two shock domain walls
with the metric  before the collision given by
\begin{equation}\label{3.2.1}
ds^{2} = L^{2}\left( - \frac{1}{z^{2}} dudv + \frac{1}{z^{2}} \phi_{1}(y, z) \delta(u) du^{2} +  \frac{1}{z^{2}} \phi_{2}(y,z)\delta(v) dv^{2} + \frac{1}{z^{2/\nu}}dy^{2} +  \frac{dz^{2}}{z^{2}}\right).
\end{equation}

The trapped surface formed in the wall-on-wall collision  obeys the equation
\begin{equation}\label{3.2.1b}
\frac{\partial^{2} \phi(z)}{\partial z^{2}}  -  \left(1 + \frac{1}{\nu}\right)\frac{1}{z} \frac{\partial \phi(z)}{\partial z} = -16\pi G_{4} E^{*}z^{1+1/\nu}\delta(z  - z_{*}).
\end{equation}

\begin{figure}[tbp]
\center{
 \includegraphics[scale=0.4]{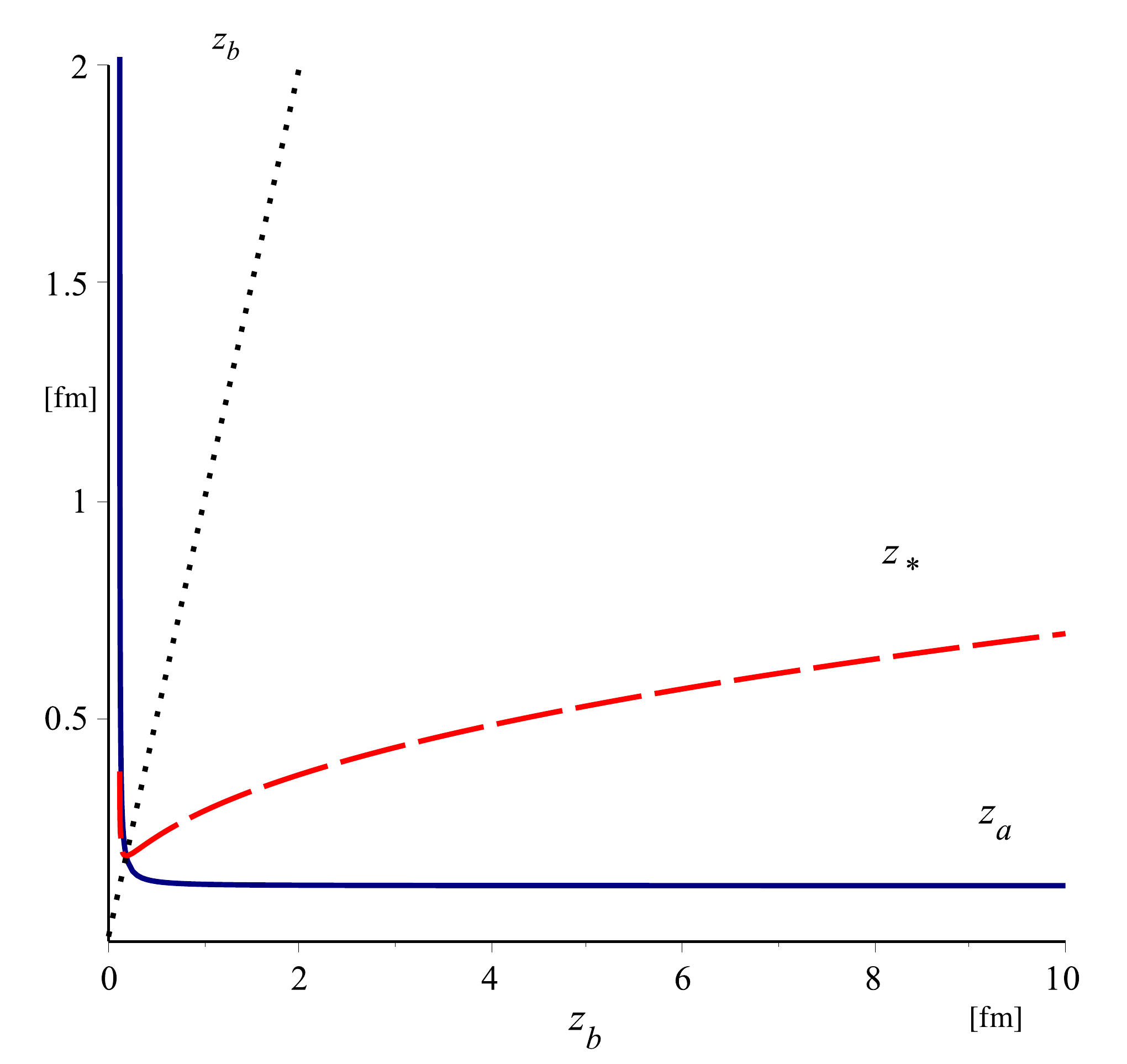}
 \caption{\label{fig:i} \textbf{The solution to the system of equations (\ref{3.2.3a}) for the given $z_{b}$ and $\nu=2$.}}}
\end{figure}

For the formation of a trapped surface the following conditions must be satisfied
\be\label{3.2.3}
(\partial_{z}\phi)|_{z = z_{a}} = 2, \quad (\partial_{z}\phi)|_{z = z_{b}} = -2, 
\ee
with the  assumption for the points $z_{a}< z_{*} <z_{b}$. The detailed form of the conditions (\ref{3.2.3}) reads
\bea\label{3.2.3a}
\frac{8\pi G_{4}E z^{1 + \frac{1}{\nu}}_{a}\left(1 - z^{\frac{1}{\nu} +2}_{b}/z^{\frac{1}{\nu} +2}_{*}\right)}{L^{3+\frac{1}{\nu}}\left(z^{\frac{1}{\nu} +2}_{b}/z^{\frac{1}{\nu} +2}_{*} - z^{\frac{1}{\nu} +2}_{a}/z^{\frac{1}{\nu} +2}_{*}\right)} & = & -1, \nn\\  \label{3.2.3c}
\frac{8\pi G_{4}E z^{1 + \frac{1}{\nu}}_{b}\left(1 - z^{\frac{1}{\nu} +2}_{b}/z^{\frac{1}{\nu} +2}_{*}\right)}{L^{3+\frac{1}{\nu}}\left(z^{\frac{1}{\nu} +2}_{b}/z^{\frac{1}{\nu} +2}_{*} - z^{\frac{1}{\nu} +2}_{a}/z^{\frac{1}{\nu} +2}_{*}\right)} &=&1.
\eea
One can derive the following relations between the collision and boundary points from eqs.(\ref{3.2.3c})  for some fixed $z_{b}$
\be\label{3.2.3b}
z_{a} = \left( \frac{z^{1+ \frac{1}{\nu}}_{b}}{- 1 +z^{1+ \frac{1}{\nu}}_{b}C}\right)^{\frac{\nu}{\nu +1 }} , \quad z_{*} =\left( \frac{z^{1+\frac{1}{\nu}}_{a}z^{1+\frac{1}{\nu}}_{b}(z_{a} + z_{b})}{z^{1+ \frac{1}{\nu}}_{a} + z^{1+ \frac{1}{\nu}}_{b}}\right)^{\frac{\nu}{2\nu +1}},
\ee
where $C = \displaystyle{ \frac{8\pi G_{4}E}{L^{\frac{1}{\nu} + 3}}}$. The solution to this system for the dynamical exponent $\nu =2$ is shown in Fig.~\ref{fig:i}.

The area of the trapped surface can be calculated using the following relation
\begin{equation}\label{3.2.4}
S = \frac{1}{2G_{4}}\int_{C}\sqrt{\textrm{det}|g_{Lif_{2}}|}dz dy,
\end{equation}
where the metric determinant of the two-dimensional Lifshitz metric (\ref{3.5b}) reads
\be \label{3.2.5}
\textrm{det}|g_{Lif_{2}}| = z^{-2(1+ \frac{1}{\nu})}.
\ee
Owing to (\ref{3.2.5}) one obtains the following result for the relative entropy
\be\label{3.2.5a}
s = \frac{S_{\textrm{trap}}}{\int dy} = \frac{\nu}{2 G_{4}}\left(\frac{1}{z^{1/\nu}_{a}} - \frac{1}{z^{1/\nu}_{b}} \right).
\ee
For large values  of $z_{b}$ we get the following approximation  
\bea\label{3.2.6}
s(C,z_{b}) &= & \left(\frac{C}{2}\right)^{\frac{1}{1+\nu}}-\left(\frac{1}{z_b}\right)^{\frac{1}{\nu}}-\frac{2}{(\nu+1)C}
\left(\frac{C}{2}\right)^{\frac{1}{1+\nu}}\,\left(\frac{1}{z_b}\right)^{\frac{1+\nu}{\nu}} + \quad \ldots  \quad.
\eea

Thus, the relative area $s$ of the trapped surface takes the maximum value at infinite $z_{b}$
\begin{equation}\label{3.2.7}
 s|_{z_{b}\rightarrow \infty} = \frac{\nu}{2G_{4}}(8\pi G_{4})^{1/(1 +\nu)}E^{1/(1 +\nu)}.
\end{equation}



\section{Shock waves in $4+1$-dimensional Lifshitz-like spacetimes} \label{Sect:3}

In this section, we construct five-dimensional Lifshitz-like shock waves. As in the previous section, we work with the Lifshitz-like metric of the form  (\ref{1.3}).
In the five-dimensional spacetime, one have two cases: $p=1, q =2$ and $p=2, q =1$.

\subsection{Type {\bf(1,2)} 5d Lifshitz shock wave} \label{Sect:3.1}

Consider a five-dimensional Lifshitz-like metric given by (\ref{1.3}) with $p = 1$, $q = 2$ written in terms of the coordinate $z = r^{-\nu}$
\begin{equation}\label{4.1.1a}
ds^{2} =L^{2}\left[\frac{\left(-dt^{2} + dx^{2}\right)}{z^{2}} + \frac{\left(dy^{2}_{1} +dy^{2}_{2}\right)}{z^{2/\nu}} +  \frac{d z^{2}}{z^{2}}\right].
\end{equation}
As in the four-dimensional case  the spacetime (\ref{4.1.1a}) for $\nu = 1$ represents the Poincare patch of anti-de Sitter space.

To construct a shock wave solution, it is convenient to represent the metric in light-cone coordinates
\begin{equation}\label{4.1.3}
ds^{2} =L^{2} \left[- \frac{dudv}{z^{2}} + \frac{\left(dy^{2}_{1} +dy^{2}_{2}\right)}{z^{2/\nu}} +  \frac{dz^{2}}{z^{2}}\right].
\end{equation}
The shock wave moving in the $v$-direction is given by
\begin{equation}\label{4.1.4a}
ds^{2} = L^{2}\left[\frac{\phi(y_{1},y_{2},z)\delta(u) }{z^{2}}du^{2} - \frac{dudv}{z^{2}} + \frac{\left(dy^{2}_{1} +dy^{2}_{2}\right)}{z^{2/\nu}} +  \frac{dz^{2}}{z^{2}}\right]
\end{equation}
with the profile function $\phi(y_{1},y_{2},z)$ satisfying the following equation
\begin{equation}\label{4.1.13}
\left[\Box_{Lif_{3}} - \frac{1}{L^{2}}\left(1 + \frac{2}{\nu}\right)\right]\frac{\phi(y_{1},y_{2},z)}{z} = -2zt_{uu}.
\end{equation}
Eq. (\ref{4.1.13}) can be obtained from the $uu$-component of the Einstein equations. Expressions for $R_{ij}$, $R$, the cosmological constant and the derivation of (\ref{4.1.13}) are presented in Appendix~\ref{App:B1}.

The Laplace-Beltrami operator $\Box_{Lif_{3}}$  has the following form
\begin{equation}\label{4.1.6a}
\Box_{Lif_{3}} =  \frac{1}{\nu L^{2}}\left(z^{2}\nu\frac{\partial^{2}}{\partial z^{2}} + \nu z\frac{\partial}{\partial z} - 2z\frac{\partial}{\partial z} + z^{2/\nu}\nu \frac{\partial^{2}}{\partial y^{2}_{1}} + \nu z^{2/\nu}\frac{\partial^{2}}{\partial y^{2}_{2}}\right)
\end{equation}
and acts on the profile function $\displaystyle{\frac{\phi(y_{1},y_{2},z)}{z}}$ as
\begin{eqnarray}\label{4.1.6b}
\Box_{Lif_{3}}\left[\frac{\phi(y_{1},y_{2},z)}{z}\right] = \frac{1}{\nu L^{2}} \Bigl[z\nu \frac{\partial^{2} \phi (y_{1},y_{2},z)}{\partial z^{2}} + z^{-1+ \frac{2}{\nu}}\nu\frac{\partial^{2}\phi(y_{1},y_{2},z)}{\partial y^{2}_{1}} + \nonumber \\ 
z^{-1+ \frac{2}{\nu}}\nu \frac{\partial^{2} \phi(y_{1},y_{2},z)}{\partial y^{2}_{2}}  -  \nu \frac{\partial \phi (y_{1},y_{2},z)}{\partial z} - 2\frac{\partial \phi(y_{1},y_{2},z)}{\partial z} + \nonumber \\ \frac{\nu}{z} \phi(y_{1},y_{2},z) + \frac{2}{z}\phi(y_{1},y_{2},z)\Bigr].
\end{eqnarray}
This operator is defined on the three-dimensional Lifshitz space with the metric
\begin{equation}\label{4.1.5b}
ds^{2} = L^{2}\left[\frac{\left(dy^{2}_{1} +dy^{2}_{2}\right)}{z^{2/\nu}} +  \frac{dz^{2}}{z^{2}}\right],
\end{equation}
owning the following Killing vectors
\begin{eqnarray}\label{4.1.5c}
 \zeta_{1} =  \frac{\partial}{\partial y_{1}}, \quad  \zeta_{2} = \frac{\partial}{\partial y_{2}}, \quad \xi = -y_{2}\frac{\partial}{\partial y_{1}} + y_{1}\frac{\partial}{\partial y_{2}}, \nonumber\\
\eta = \frac{y_{1}}{\nu}\frac{\partial}{\partial y_{1}} + \frac{y_{2}}{\nu}\frac{\partial}{\partial y_{2}} + z\frac{\partial}{\partial z}, \quad \eta_{1}=  -\frac{z^{2/\nu}\nu^{2} - y_{1}^{2} + y_{2}^{2} }{2\nu}\frac{\partial}{\partial y_{1}} + \frac{y_{1}y_{2}}{\nu}\frac{\partial}{\partial y_{2}} + y_{1}z \frac{\partial}{\partial z}, \nonumber\\ 
\eta_{2} = -\frac{y_{1}^{2} - y_{2}^{2} +z^{2/\nu}\nu^{2}}{2\nu}\frac{\partial }{\partial y_{2}} + zy_{2}\frac{\partial}{\partial z} + y_{1}y_{2}\frac{\partial}{\partial y_{1}} .
\end{eqnarray}

Putting the dynamical exponent $\nu =1$, eq. (\ref{4.1.13}) comes to the well-known equation for the $5d$ $AdS$-shock wave
\begin{equation}\label{AdS5}
\left[\Box_{H_{3}} - \frac{3}{L^{2}}\right]\frac{\phi(y_{1},y_{2},z)}{z} = -2zt_{uu}.
\end{equation}

\subsubsection{Domain-wall}
The equation  for the domain-wall profile in the five-dimensional Lifshitz-like space is
\begin{equation}\label{4.2.1}
\left[\Box_{Lif_{3}} - \frac{1}{L^{2}}\left(1 + \frac{2}{\nu}\right)\right]\frac{\phi(z)}{z} = -16\pi G_{5}z J_{uu},
\end{equation}
which can be represented in the form
\begin{equation}\label{4.2.1b}
\frac{\partial^{2} \phi(z)}{\partial z^{2}}  -  \left(1 + \frac{2}{\nu}\right)\frac{1}{z} \frac{\partial \phi(z)}{\partial z} = -16\pi G_{5} J_{uu},
\end{equation}
with the source given by  
\begin{equation}\label{4.2.1c}
J_{uu} = E \left(\frac{z}{L}\right)^{1+2/\nu}\delta(z  - z_{*}).
\end{equation}

The solution to the domain wave profile has the similar form to the 4-dimensional one and  reads
\begin{equation}\label{4.2.3a}
\phi =\phi_{a}\Theta(z_{*} - z) +\phi_{b}\Theta(z - z_{*}),
\end{equation}
where the profile functions are 
\begin{eqnarray}\label{4.2.4}
\phi_{a}(z) = C_{0}z_{a}z_{b}\left(\frac{z^{2(\nu+1)/\nu}_{*}}{z^{2(\nu+1)/\nu}_{b}} - 1\right)\left(\frac{z^{2(\nu+1)/\nu}}{z^{2(\nu+1)/\nu}_{a}} - 1\right),\\
\phi_{b}(z) = C_{0}z_{a}z_{b}\left(\frac{z^{2(\nu+1)/\nu}_{*}}{z^{2(\nu+1)/\nu}_{a}} - 1\right)\left(\frac{z^{2(\nu+1)/\nu}}{z^{2(\nu+1)/\nu}_{b}} - 1\right),\\ \label{4.2.4c}
C_{0} =  -\frac{8\nu\pi G_{5} E z^{1+2/\nu}_{a}z^{1+2/\nu}_{b}}{(\nu+1)L^{3+\frac{2}{\nu}}(z^{2(\nu+1)/\nu}_{b} - z^{2(\nu+1)/\nu}_{a})}.
\end{eqnarray}

\begin{figure}[tbp]
\center{
\includegraphics[scale=0.4]{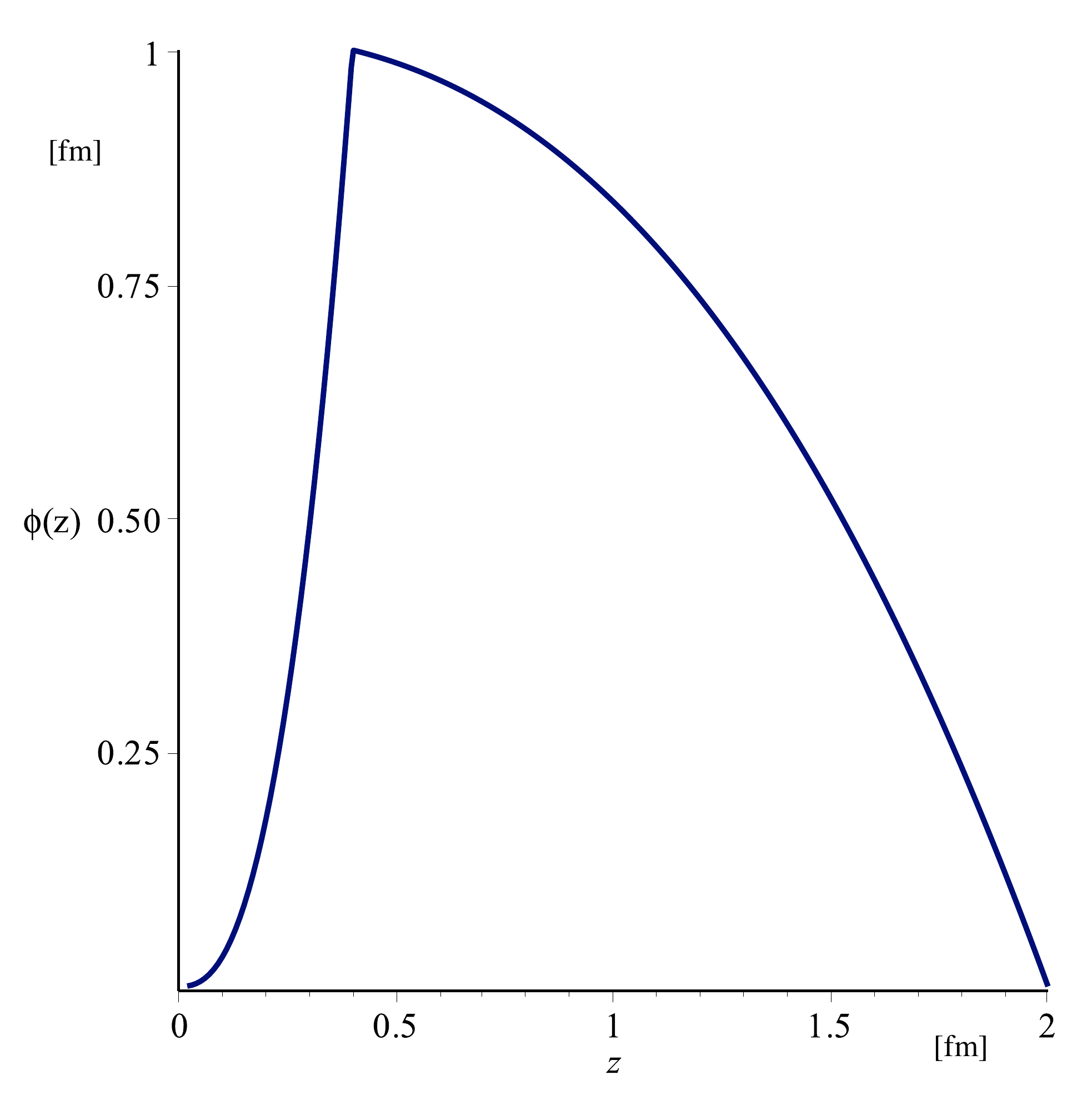}
\caption{\label{fig:2}\textbf{The domain wall profile $\phi(z)$ in the 5d Lifshitz-like spacetime {\bf(1,2)} with $\nu = 4$.}}}
\end{figure}

The profile is presented in Fig.~\ref{fig:2}.

\subsubsection{Colliding shock waves}

The line element for colliding shocks in the $5d$ Lifshitz-like background can be represented as:
\begin{eqnarray}\label{4.2.3}
ds^{2} = L^{2}\Bigl[ - \frac{1}{z^{2}} dudv + \frac{1}{z^{2}} \phi_{1}(y_{1},y_{2}, z) \delta(u) du^{2} +  \frac{1}{z^{2}} \phi_{2}(y_{1},y_{2},z)\delta(v) dv^{2} + \nn\\ \frac{1}{z^{2/\nu}}\left(dy^{2}_{1} +dy^{2}_{2}\right) +  \frac{dz^{2}}{z^{2}}\Bigr].
\end{eqnarray}

The trapped surface obeys the conditions for the formation 
\bea\label{4.3.5a}
\frac{8\pi G_{5}E z^{1 + \frac{2}{\nu}}_{a}\left(1 - z^{\frac{2}{\nu} +2}_{b}/z^{\frac{2}{\nu} +2}_{*}\right)}{L^{3+\frac{2}{\nu}}\left(z^{\frac{2}{\nu} +2}_{b}/z^{\frac{2}{\nu} +2}_{*} - z^{\frac{2}{\nu} +2}_{a}/z^{\frac{2}{\nu} +2}_{*}\right)} & = & -1, \nn\\  \label{4.3.5c}
\frac{8\pi G_{5}E z^{1 + \frac{2}{\nu}}_{b}\left(1 - z^{\frac{2}{\nu} +2}_{b}/z^{\frac{2}{\nu} +2}_{*}\right)}{L^{3+\frac{2}{\nu}}\left(z^{\frac{2}{\nu} +2}_{b}/z^{\frac{2}{\nu} +2}_{*} - z^{\frac{2}{\nu} +2}_{a}/z^{\frac{2}{\nu} +2}_{*}\right)} &=&1.
\eea
Owing to (\ref{4.3.5a}) we obtain the following relations between the collision and boundary points 
\be\label{4.3.5b}
z_{a} = \left( \frac{z^{1+ \frac{2}{\nu}}_{b}}{- 1 +z^{1+ \frac{2}{\nu}}_{b}C}\right)^{\frac{\nu}{\nu + 2}} , \quad z_{*} =\left( \frac{z^{1+\frac{2}{\nu}}_{a}z^{1+\frac{2}{\nu}}_{b}(z_{a} + z_{b})}{z^{1+ \frac{2}{\nu}}_{a} + z^{1+ \frac{2}{\nu}}_{b}}\right)^{\frac{\nu}{2\nu + 2}},
\ee
where $C = \displaystyle{ \frac{8\pi G_{5}E}{L^{\frac{2}{\nu} + 3}}}$.

One can calculate the area of the trapped surface using the relation
\begin{equation}\label{4.3.6}
S = \frac{1}{2G_{5}}\int_{C}\sqrt{\textrm{det}|g_{Lif_{3}}|}dz dy_{1}dy_{2},
\end{equation}
with  the  determinant $\textrm{det}|g_{Lif_{3}}|$ of the three-dimensional Lifshitz metric (\ref{4.1.5b})
\be\label{4.3.6c}
\textrm{det}|g_{Lif_{3}}|  = z^{-(1+\frac{2}{\nu})}.
\ee

 Thus, we  have 
\begin{equation}\label{4.3.6a}
s =\frac{S_{\textrm{trap}}}{\int dy_{1}dy_{2}} = \frac{\nu}{4G_{5}}\left(\frac{1}{(z_{a})^{2/\nu}} - \frac{1}{(z_{b})^{2/\nu}}\right).
\end{equation}

In the limit $z_{b}\rightarrow \infty$ the approximation for the entropy  can be  represented as
\bea\label{4.3.7}
s(C,z_{b}) &= & \left(\frac{C}{2}\right)^{\frac{2}{2+\nu}}-\left(\frac{1}{z_b}\right)^{\frac{2}{\nu}}-\frac{2}{(\nu+2)}
\left(\frac{2}{C}\right)^{\frac{2}{2+\nu}}\,\left(\frac{1}{z_b}\right)^{\frac{2+\nu}{\nu}} + \quad \ldots \quad.
\eea

The maximum value of the relative area $s$ at infinite $z_{b}$ takes the form 
\begin{equation}\label{4.3.6b}
 s|_{z_{b}\rightarrow \infty} = \frac{\nu}{4G_{5}} (8\pi G_{5})^{2/(\nu+2)}E^{2/(\nu+2)}.
\end{equation}

We can see that relations (\ref{4.2.4})-(\ref{4.2.4c})  and (\ref{4.3.6b}) are similar to those for the 5-dimensional AdS background deformed 
by the power-law factor $b =\displaystyle{\frac{L}{z^a}}$  from \cite{APP}  with the profile equation
\begin{equation}\label{4.2.1d}
\frac{\partial^{2} \phi(z)}{\partial z^{2}}  - \frac{3a}{z}\frac{\partial \phi(z)}{\partial z} = -16\pi G_{5}\left(\frac{z}{L}\right)^{3a}E^{*}\delta(z - z_{*}).
\end{equation}
One can conclude that  the following relation between the constant $a$ and the Lifshitz exponent $\nu$ takes place
\be\label{4.3.8}
1 + \frac{2}{\nu} = 3a.
\ee
In \cite{APP} it is shown that the $b$-factor with $a = 1/2$ gives rise to the value of multiplicity, which is the most compatible to the experimental data. 
Thus, one should consider wall-on-wall-collisions in the Lifshitz-like spacetime with $\nu =4$.

\subsection{Type {\bf(2,1)} 5d Lifshitz shock wave}\label{Sect:3.2}
Now we turn to constructing shocks in a five-dimensional Lifshitz-like background with $p=2$, $q=1$
\begin{equation}\label{4.4.1}
ds^{2} = L^{2}\left( \frac{\left(-dt^{2} + dx^{2}_{1} + dx^{2}_{2}\right)}{z^{2}} + \frac{dy^{2}}{z^{2/\nu}} +  \frac{dz^{2}}{z^{2}}\right).
\end{equation}
The metric (\ref{4.4.1}) can be rewritten in the following form
\begin{equation}\label{4.4.3}
ds^{2} = L^{2}\left(- \frac{dudv}{z^{2}} + \frac{dx^{2}_{2}}{z^{2}} + \frac{dy^{2}_{2}}{z^{2/\nu}} +  \frac{dz^{2}}{z^{2}}\right),
\end{equation}
where $du = dt - dx_{1}$ and $dv = dt + dx_{1}$. 

The metric of the shock wave in the background (\ref{4.4.1}) is given by
\begin{equation}\label{4.4.4}
ds^{2} = L^{2}\left(\frac{\phi(x_{2},y_{2},z)\delta(u) }{z^{2}}du^{2} - \frac{1}{z^{2}} dudv + \frac{dx^{2}_{2}}{z^{2}} + \frac{dy^{2}_{2}}{z^{2/\nu}} +  \frac{dz^{2}}{z^{2}}\right).
\end{equation}
The equation for the shock wave profile reads
\begin{equation}\label{4.4.10}
\left[\Box_{Lif_{3}} -  \frac{1}{L^{2}}\left(2 + \frac{1}{\nu}\right)\right]\frac{\phi(x_{2},y_{2},z)}{z} = -2zt_{uu},
\end{equation}
which follows from the Einstein equations (see, Appendix ~\ref{App:B2}).

In eq. (\ref{4.4.10}) the Laplace-Beltrami operator $\Box_{Lif_{3}}$ is given by
\begin{equation}\label{4.4.6a}
\Box_{Lif_{3}} =  \frac{1}{\nu L^{2}}\left(z^{2}\nu\frac{\partial^{2}}{\partial z^{2}} - z\frac{\partial}{\partial z} + z^{2}\nu \frac{\partial^{2}}{\partial x^{2}_{2}} + \nu z^{2/\nu}\frac{\partial^{2}}{\partial y^{2}_{2}}\right).
\end{equation}
and acts on the profile function $\displaystyle{\frac{\phi(x_{2},y_{2},z)}{z}}$ as
\begin{eqnarray}\label{4.4.6b}
\Box_{Lif_{3}}\frac{\phi(x_{2},y_{2},z)}{z} = \frac{1}{\nu L^{2}}\Bigl[z \nu \frac{\partial^{2} \phi(x_{2}, y_{2}, z)}{\partial z^{2}}   + z \nu \frac{\partial^{2} \phi(x_{2}, y_{2}, z)}{\partial x^{2}_{2}} + z^{-1 + 2/\nu} \nu \frac{\partial^{2} \phi(x_{2}, y_{2}, z)}{\partial y^{2}_{2}} - \nonumber\\
- 2 \nu \frac{\partial \phi(x_{2}, y_{2}, z)}{\partial z} -  \frac{\partial \phi(x_{2}, y_{2}, z)}{\partial z} + 2 \frac{\nu}{z}\phi(x_{2},y_{2},z) + \frac{1}{z}\phi(x_{2},y_{2},z) \Bigr].\nn
\\
\end{eqnarray}
$\Box_{Lif_{3}}$ is defined on the three-dimensional Lifshitz space
\begin{equation}\label{4.4.5b}
ds^{2} = L^{2} \left(\frac{dx^{2}_{2}}{z^{2}} +\frac{dy^{2}_{2}}{ z^{2/\nu}} +  \frac{dz^{2}}{z^{2}}\right).
\end{equation}
Killing vectors related to (\ref{4.4.5b}) have the form
\begin{equation}\label{4.4.6}
\xi = x_{2}\frac{\partial}{\partial x_{2}} + \frac{y_{2}}{\nu}\frac{\partial}{\partial y_{2}} + z\frac{\partial}{\partial z}, \quad \zeta_{1} =  \frac{\partial}{\partial x_{2}}, \quad \zeta_{2} = \frac{\partial}{\partial y_{2}}.
\end{equation}

\subsubsection{Domain-wall}
The profile of a domain wall has  the dependence only on holographic coordinate $z$ and obeys the equation
\begin{equation}\label{6.1}
\left[\Box_{Lif_{3}} -  \frac{1}{L^{2}}\left(2 + \frac{1}{\nu}\right)\right]\frac{\phi(z)}{z} = -16\pi G_{5}J_{uu}.
\end{equation}
Owing to the absence of the dependence on the transversal coordinates, eq. (\ref{6.1}) can be reduced to the form
\begin{equation}\label{6.1b}
\frac{\partial^{2} \phi(z)}{\partial z^{2}}  -  \left(2 + \frac{1}{\nu}\right)\frac{1}{z} \frac{\partial \phi(z)}{\partial z} = -16\pi G_{5}E^{*} z^{2+1/\nu}\delta(z  - z_{*}).
\end{equation}
The  domain wave profile can be represented in the following form
\begin{equation}\label{6.2a}
\phi =\phi_{a}\Theta(z_{*} - z) +\phi_{b}\Theta(z - z_{*}),
\end{equation}
where the profile functions are 
\begin{eqnarray}\label{6.2}
\phi_{a}(z) = C_{0}z_{a}z_{b}\left(\frac{z^{(3\nu+1)/\nu}_{*}}{z^{(3\nu+1)/\nu}_{b}} - 1\right)\left(\frac{z^{(3\nu+1)/\nu}}{z^{(3\nu+1)/\nu}_{a}} - 1\right),\\
\phi_{b}(z) = C_{0}z_{a}z_{b}\left(\frac{z^{(3\nu+1)/\nu}_{*}}{z^{(3\nu+1)/\nu}_{a}} - 1\right)\left(\frac{z^{(3\nu+1)/\nu}}{z^{(3\nu+1)/\nu}_{b}} - 1\right),\\ \label{6.2b}
C_{0} =  -\frac{16\nu\pi G_{5} E z^{2+1/\nu}_{a}z^{2+1/\nu}_{b}}{(3\nu+1)L^{4+\frac{1}{\nu}}(z^{(3\nu+1)/\nu}_{b} - z^{(3\nu+1)/\nu}_{a})}.
\end{eqnarray}

Comparing the equations (\ref{6.1b}) and (\ref{4.2.1d}), one obtains the following correspondence 
\begin{equation}\label{6.10}
3a = 2 + \frac{1}{\nu}.
\end{equation}
For $\nu = -2$ the solution for the domain wall profile (\ref{6.2a})-(\ref{6.2b}) yields the same results as for the deformed AdS case \cite{APP} with $a = 1/2$. 
However,  it is shown in \cite{HK} that Lifshitz geometries with negative dynamical constants do not satisfy the null energy condition.

\subsubsection{Colliding shocks}
The ansatz of the metric for colliding domain walls in the Lifshitz-like background (\ref{4.4.1}) is
\begin{eqnarray}\label{6.3}
ds^{2} = L^{2}\Bigl[ - \frac{1}{z^{2}} dudv + \frac{1}{z^{2}} \phi_{1}(y_{1},y_{2}, z) \delta(u) du^{2} +  \frac{1}{z^{2}} \phi_{2}(y_{1},y_{2},z)\delta(v) dv^{2} + \nn\\ \frac{dy^{2}_{1}}{z^{2/\nu}} + \frac{dy^{2}_{2}}{z^{2}} +  \frac{dz^{2}}{z^{2}}\Bigr].
\end{eqnarray}
The trapped surface is formed if  the following conditions are satisfied
\bea\label{6.4}
\frac{8\pi G_{5}E z^{2 + \frac{1}{\nu}}_{a}\left(1 - z^{\frac{1}{\nu} +3}_{b}/z^{\frac{1}{\nu} +3}_{*}\right)}{L^{4+\frac{1}{\nu}}\left(z^{\frac{1}{\nu} + 3}_{b}/z^{\frac{1}{\nu} +3}_{*} - z^{\frac{1}{\nu} +3}_{a}/z^{\frac{1}{\nu} + 3}_{*}\right)} & = & -1, \nn\\  \label{6.3.5c}
\frac{8\pi G_{5}E z^{2 + \frac{1}{\nu}}_{b}\left(1 - z^{\frac{1}{\nu} + 3}_{b}/z^{\frac{1}{\nu} +3}_{*}\right)}{L^{4+\frac{1}{\nu}}\left(z^{\frac{1}{\nu} +3}_{b}/z^{\frac{1}{\nu} +3}_{*} - z^{\frac{1}{\nu} +3}_{a}/z^{\frac{1}{\nu} + 3}_{*}\right)} &=&1.
\eea
The  relations between the collision and boundary points read
\be\label{6.5}
z_{a} = \left( \frac{z^{2+ \frac{1}{\nu}}_{b}}{- 1 +z^{2+ \frac{1}{\nu}}_{b}C}\right)^{\frac{\nu}{2\nu + 1}} , \quad z_{*} =\left( \frac{z^{2+\frac{1}{\nu}}_{a}z^{2+\frac{1}{\nu}}_{b}(z_{a} + z_{b})}{z^{2+ \frac{1}{\nu}}_{a} + z^{2+ \frac{1}{\nu}}_{b}}\right)^{\frac{\nu}{3\nu + 1}},
\ee
where $C = \displaystyle{ \frac{8\pi G_{5}E}{L^{\frac{1}{\nu} + 4}}}$.

The metric determinant for the $3d$ Lifshitz space (\ref{4.4.5b})  is
\be \label{6.6}
\textrm{det}|g_{Lif_{3}}| = z^{-(4 +\frac{2}{\nu})}.
\ee

The relative entropy is given by
\begin{equation}\label{6.7}
s =\frac{S_{\textrm{trap}}}{\int dy_{1}dy_{2}} = \frac{\nu}{2G_{5}(\nu+1)}\left(\frac{1}{z^{1+\frac{1}{\nu}}_{a}} - \frac{1}{z^{1+\frac{1}{\nu}}_{b}}\right).
\end{equation}

Now we can write down the approximation for the entropy in the limit $z_{b}\rightarrow \infty$
\bea\label{6.8}
s(C,z_{b}) &= & \left(\frac{C}{2}\right)^{\frac{\nu + 1}{2\nu+1}}-\left(\frac{1}{z_b}\right)^{1 +\frac{1}{\nu}}- \frac{\nu +1}{2\nu +2}
\left( \frac{2}{C}\right)^{\frac{\nu}{2\nu+1}}\,\left(\frac{1}{z_b}\right)^{\frac{2\nu + 1}{\nu}}+\quad \ldots \quad .
\eea

The relative area $s$ of the trapped surface, the maximum value at infinite $z_{b}$
\begin{equation}\label{6.9}
 s|_{z_{b}\rightarrow \infty} = \frac{\nu}{2G_{5}(\nu+1)}(8\pi G_{5})^{\frac{\nu + 1}{2\nu+1}}E^{\frac{\nu + 1}{2\nu+1}}.
\end{equation}

\section{Shock waves for brane systems}\label{Sect:4}
In this section, we will study two brane systems defined on a product of manifolds, which contains a Lifshitz-like metric. D3-D7 intersection that we will consider is the IR part of the zero-temperature solution from \cite{ALT}-\cite{MT2}.

\subsection{Intersecting D3-D7 branes}

First, we briefly review the related IIB supergravity model. The action can be written in the following form
\begin{eqnarray}\label{7.1}
S =\int  d^{10}x \sqrt{|g|} \Bigl\{ e^{-2\phi}(R + 4 \partial_{M}\varphi \partial^{M}\varphi - \frac{1}{2} H_{3} \wedge \star H_{3}) - \frac{1}{2}F_{1}\wedge \star F_{1} - \nn\\
 \frac{1}{2}\tilde{F}_{3}\wedge \tilde{F}_{3} - \frac{1}{4}\tilde{F}_{5}\wedge \star \tilde{F}_{5} \Bigr\}- \frac{1}{2} \int C_{4}\wedge H_{3} \wedge F_{3},
\end{eqnarray}
$\varphi$ is the dilaton field,
the field strengths have the following relations $F_{1}= d \chi $, $\tilde{F}_{3} = F_{3} - \chi H_{3}$, 
$\tilde{F}_{5} = F_{5}  - \frac{1}{2}C_{2}\wedge H_{3} + \frac{1}{2} B_{2} \wedge F_{3}$ and $\chi$ is the axion field.

We consider a $D3-D7$-brane system defined on a ten-dimension manifold $\mathcal{M}$, which can be factorized as
\begin{equation}\label{7.1a}
\mathcal{M} = M_{5}\times X_{5},
\end{equation}
where $X_{5}$ is an Einstein manifold and $M_{5}$ is a manifold with anisotropy.
According to \cite{ALT} the Lifshitz-type metric can appear in  a solution which describes  intersecting $D3$ and $D7$ branes.
The metric of the $D3-D7$-intersection is given by
\begin{equation}\label{7.2a}
ds^{2}_{E} = \tilde{R}^{2}\left[r^{3}(-dt^{2} + dx^{2} + dy^{2}) + r^{2}dw^{2} + \frac{dr^{2}}{r^{2}}\right] + R^{2}ds^{2}_{X_{5}},
\end{equation}
where the radii 
\begin{equation}\label{7.3}
R^{2} = \frac{12}{11}\tilde{R}^{2}  = \frac{\alpha}{2}.
\end{equation}
The configuration is given in Fig.~\ref{fig:3}. 
Redefining the coordinate $r$ as $r \equiv \rho^{2/3}$ one obtains

\begin{equation}\label{7.4}
ds^{2} = \tilde{R}^{2}\left[\rho^{2}\left(-dt^{2} + dx^{2}+dy^{2}\right) + \rho^{4/3}dw^{2} + \frac{d\rho^{2}}{\rho^{2}}\right] + R^{2}ds^{2}_{X_{5}}.
\end{equation}

The spacetime (\ref{7.4}) is invariant under the scaling
\begin{equation}\label{7.5}
(t,x,y,w,\rho) \rightarrow\left(\lambda t,\lambda x, \lambda y, \lambda^{2/3}w, \frac{\rho}{\lambda} \right)
\end{equation}
and coincides with (\ref{1.3}) for $p  = 2$, $q = 1$ and the critical exponent  $\nu = 3/2$.

\begin{figure}[!h]
\begin{equation}\label{D3-D7}
\begin{array}{r|cccc|c|ccccccl}
\,\, \mbox{$\mathcal{M}_4\times S^1\times X_5$}\,\,\, & t     & x
& y & r  & w &  s_1   & s_2   & s_3   & s_4 & s_5 &\, \nonumber\\
\hline  \,\, \mbox{D3}\,\,\,& \times &   \times  &  \times &   & \times &   &  & &
&
&\,\nonumber\\
\,\, \mbox{D7}\,\,\,& \times &   \times  &  \times &   &  &  \times & \times & \times &
\times& \times &\,\nonumber
\end{array}
\end{equation}
\caption{\label{fig:3}D3-D7 intersection over a two-brane.}
\end{figure}


Introducing light-cone coordinates $du = dt - dx$ and $dv = dt + dx$ and rewriting (\ref{7.4}) in terms of the holographic coordinate $z = 1/r$, one comes
\begin{equation}\label{8.6b}
ds^{2} = \tilde{R}^{2}\left[- \frac{dudv}{z^{2}} + \frac{dy^{2}}{z^{2}} + \frac{dw^{2}}{z^{4/3}} + \frac{dz^{2}}{z^{2}}\right] + R^{2}ds^{2}_{X_{5}}.
\end{equation}

We consider a shock wave which is defined on a space including the common brane worldvolume, the relative transverse space to the $D7$-brane and the one-dimensional part of the total transverse space. 
For simplicity, we restrict ourselves to the case $\phi = \phi(z)$ that yields  the  domain wall geometry. Thus, the metric of the shock is given by
\begin{equation}\label{8.6d}
ds^{2} = \tilde{R}^{2}\left[\frac{\phi(z)}{z^{2}} \delta(u)du^{2} - \frac{dudv}{z^{2}} + \frac{dy^{2}}{z^{2}} + \frac{dw^{2}}{z^{4/3}} + \frac{d z^{2}}{z^{2}}\right]
\end{equation}
and is invariant under the Lifshitz-like transformations (\ref{7.5}). 
Now it is easy to see that the geometry (\ref{8.6d}) is a particular case of the shock wave in the Lifshitz metric type $(\bf{2,1})$ (\ref{4.4.4})  with $L = \tilde{R}$.
The equation for the profile can be represented in the form
\begin{equation}\label{8.7c}
\frac{\partial^{2} \phi(z)}{\partial z^{2}}  -  \frac{8}{3z} \frac{\partial \phi(z)}{\partial z} = -16\pi G_{5} E^{*}z^{8/3}\delta(z  - z_{*})
\end{equation}
and has the  solution (\ref{6.2a})-(\ref{6.2b}) with putting $\nu = 3/2$.

\begin{figure}[tbp]
\center{
 \includegraphics[scale=0.4]{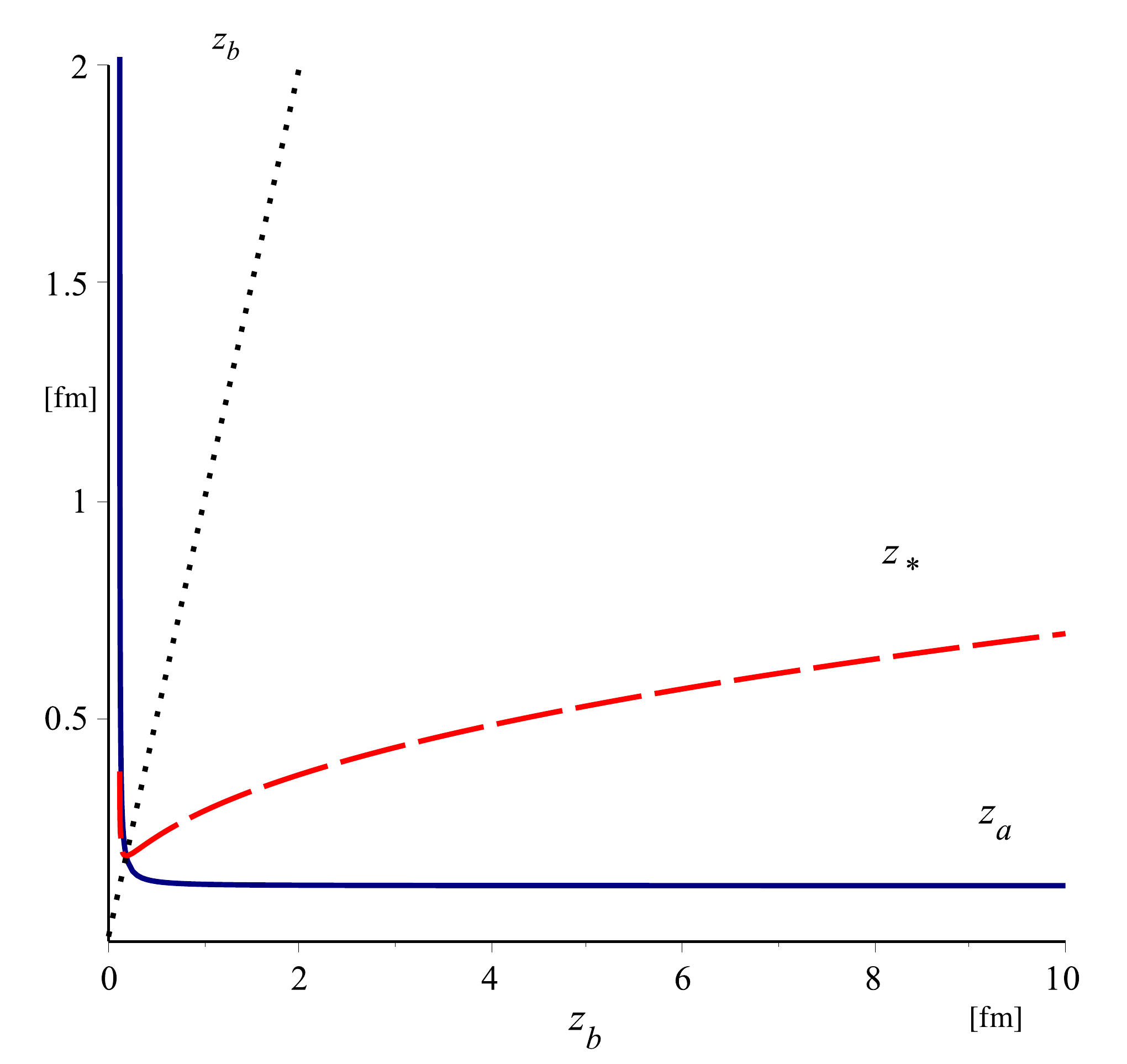}
 \caption{\label{fig:4}\textbf{The solution for the boundary and collision points (\ref{8.9a}).}}}
\end{figure}

The ansatz for the metric  before the collision of two domain walls in the background (\ref{8.6b}) has the similar form to (\ref{6.3})
\bea\label{8.9}
ds^{2} = \tilde{R}^{2} \left[ - \frac{1}{z^{2}} dudv + \frac{1}{z^{2}} \phi_{1}(z) \delta(u) du^{2} +  \frac{1}{z^{2}} \phi_{2}(z)\delta(v) dv^{2} + 
\frac{dy^{2}}{z^{2}} + \frac{dw^{2}}{z^{4/3}} +  \frac{dz^{2}}{z^{2}}\right].
\eea
The trapped surface can be formed if the following boundary conditions are satisfied
\bea\label{8.9a}
\frac{8\pi G_{5}Ez^{8/3}_{a}\left(1- \frac{z^{11/3}_{b}}{z^{11/3}_{*}}\right)}{\tilde{R}^{14/3}\left(\frac{z^{11/3}_{b}}{z^{11/3}_{*}} - \frac{z^{11/3}_{a}}{z^{11/3}_{*}} \right)} = - 1, \quad \frac{8\pi G_{5}Ez^{8/3}_{b}\left(1- \frac{z^{11/3}_{a}}{z^{11/3}_{*}}\right)}{\tilde{R}^{14/3}\left(\frac{z^{11/3}_{b}}{z^{11/3}_{*}} - \frac{z^{11/3}_{a}}{z^{11/3}_{*}} \right)} = 1.
\eea
The solution to (\ref{8.9a}) is given in Fig.~\ref{fig:4}. The relations for the collision and boundary points are those (\ref{6.5}) with $\nu = 3/2$.

Following the calculations (\ref{6.6})-(\ref{6.8}) one can write down  the maximum value of  the relative area $s$ of the trapped surface at infinite $z_{b}$
\begin{equation}\label{8.9d}
s|_{z_{b}\rightarrow \infty} = \frac{3\tilde{R}}{10G_{5}}\left(\frac{8\pi G_{5}}{\tilde{R}^{2}}\right)^{5/8}E^{5/8}.
\end{equation}

\subsection{Intersecting D4-D6 branes}

The action in  type IIA supergravity has the following form

\begin{eqnarray}\label{8.2.1}
S_{IIA} = \int d^{10}x \sqrt{|g|}\Bigl\{e^{-2\varphi}\left[R[g]+ 4\partial_{\mu}\varphi\partial^{\mu}\varphi - \frac{1}{2(3!)}|F_{(3)}|^{2}\right] - 
\frac{1}{2(2!)}|F_{(2)}|^{2} - \nn \\ \frac{1}{2(4!)}|\tilde{F}_{(4)}|^{2} \Bigr\} - 
 \frac{1}{2}\int A_{2}\wedge F_{(4)}\wedge F_{(4)},
\end{eqnarray}
where  $\varphi$ is the dilaton,  $F_{(3)} = dA_{2}$ is the field strength of the NS-NS two form,
 $F_{(2)} = dA_{1}$  is the field strength of the R-R 1-form, $F_{(4)} = dA_{3}$,  $\tilde{F}_{(4)} = dA_{3} + F_{(3)}\wedge A_{1}$ are the Ramond-Ramond field strengths.

The D4-D6 brane system with a Lifshitz-like scaling from \cite{ALT} has the following  form
\begin{eqnarray}\label{8.2.2}
ds^{2} = \tilde{R}^{2}\left[\rho^{7/3}\left(-dt^{2} + dx^{2}_{1} + dx^{2}_{2}\right) + \frac{d\rho^{2}}{\rho^{5/3}} + \rho^{5/3}\left(dy^{2}_{1} + dy^{2}_{2}\right)\right] + R^{2}\rho^{1/3}ds^{2}_{X_{4}}.
\end{eqnarray}
This configuration is presented in Fig.~\ref{fig:5}.

\begin{figure}[!h]
\begin{equation}\label{D4-D6}
\begin{array}{r|cccc|cc|cccccl}
\,\, \mbox{$\mathcal{M}_4\times X_2\times X_4$}\,\,\, & t     & x_{1}
& x_{2} & \rho  & y_{1} &  y_2   & s_1   & s_2   & s_3 & s_4 &\, \nonumber\\
\hline  \,\, \mbox{D4}\,\,\,& \times &   \times  &  \times &   & \times &\times  &  & &
&
&\,\nonumber\\
\,\, \mbox{D6}\,\,\,& \times &   \times  &  \times &   &  &   & \times & \times &
\times& \times &\,\nonumber
\end{array}
\end{equation}
\caption{\label{fig:5}D4-D6 intersection over a two-brane.}
\end{figure}

The D4-D6 solution  (\ref{8.2.2}) has been considered in \cite{ALT} as a solution with the Lifshitz-like geometry (\ref{1.3}), but without the conformal invariance. 
However, one should note that (\ref{8.2.2}) can be taken to the form
\begin{eqnarray}\label{8.2.4}
ds^{2} =z^{-1/3}\left\{ \tilde{R}^{2}\left[\frac{\left(-dt^{2} + dx^{2}_{1} + dx^{2}_{2}\right)}{z^{2}} + \frac{dz^{2}}{z^{2}} + \frac{\left(dy^{2}_{1} + dy^{2}_{2}\right)}{z^{4/3}}\right] + R^{2}ds^{2}_{X_{4}}\right\}.\nn\\
\end{eqnarray}
One can see that (\ref{8.2.4}) is the Lifshitz solution with the hyperscaling violation \cite{CGKKM}-\cite{DHKYW}
\bea\label{hvLif}
ds^{2} = \rho^{2\theta/d}\left[\rho^{2}\left(-dt^{2} + \sum^{p}_{i =1}dx^{2}_{i} \right) + \rho^{2/\nu}\sum^{q}_{j =1}dy^{2}_{i} + \frac{d\rho^{2}}{\rho^{2}}\right]
\eea
with  parameters
\be\label{8.2.5}
\theta = \frac{2}{3}, \quad d = 4, \quad \nu = \frac{3}{2},
\ee
where $d$ is the number of the spatial coordinates ($x_{i}$ and $y_{i}$, $i =1,2$) and $\theta$ is the hyperscaling violation exponent. 
The geometry (\ref{hvLif}) possesses the scaling property
\be
t \rightarrow \lambda t, \quad x_{i} \rightarrow \lambda x_{i},\quad z \rightarrow \lambda z, \quad y_{i} \rightarrow \lambda^{2/3} y_{i}, \quad ds \rightarrow \lambda^{-1/6} ds.
\ee
We define the shock metric on the D4-D6 system in a similar way to the shock wave in the D3-D7 background (\ref{8.6d})
\be\label{8.2.4a}
ds^{2} = \frac{\tilde{R}^{2}}{z^{1/3}}\left[\frac{\phi(z)}{z^{2}}\delta(u)du^{2} - \frac{dudv}{z^{2}} + \frac{dx^{2}_{2}}{z^{2}} + \frac{dz^{2}}{z^{2}} + \frac{\left(dy^{2}_{1} + dy^{2}_{2}\right)}{z^{4/3}}\right].
\ee
The shock profile obeys the equation
\bea\label{8.2.6a}
 \frac{\partial^{2} \phi(z)}{\partial z^{2}}  - \frac{4}{z}\frac{\partial \phi(z)}{\partial z} = -16\pi G_{6}z^{4}_{*}E\delta(z -z_{*}).
\eea
 
The solution to (\ref{8.2.6a}) is given by
\begin{equation}\label{8.2.8}
\phi(z) = \phi_{a}\Theta(z_{*} - z) + \phi_{b}\Theta(z - z_{*}),
\end{equation}
where 
\begin{eqnarray}\label{8.2.8a}
\phi_{a} &=& C_{0}z_{a}z_{b}\left(\left(\frac{z_{*}}{z_{b}}\right)^{5} - 1\right)\left(\left(\frac{z}{z_{a}}\right)^{5} - 1\right), \nn\\
\phi_{b}& =& C_{0}z_{a}z_{b}\left(\left(\frac{z_{*}}{z_{a}}\right)^{5} - 1\right)\left(\left(\frac{z}{z_{b}}\right)^{5} - 1\right), 
\end{eqnarray}
and 
\be\label{8.2.8c}
C_{0} =\displaystyle{ -\frac{16\pi G_{5} E}{6\tilde{R}^{6}}\frac{z^{4}_{a}z^{4}_{b}}{(z^{5}_{b} - z^{5}_{a})}}.
\ee

\begin{figure}[tbp]
\center{
 \includegraphics[scale=0.4]{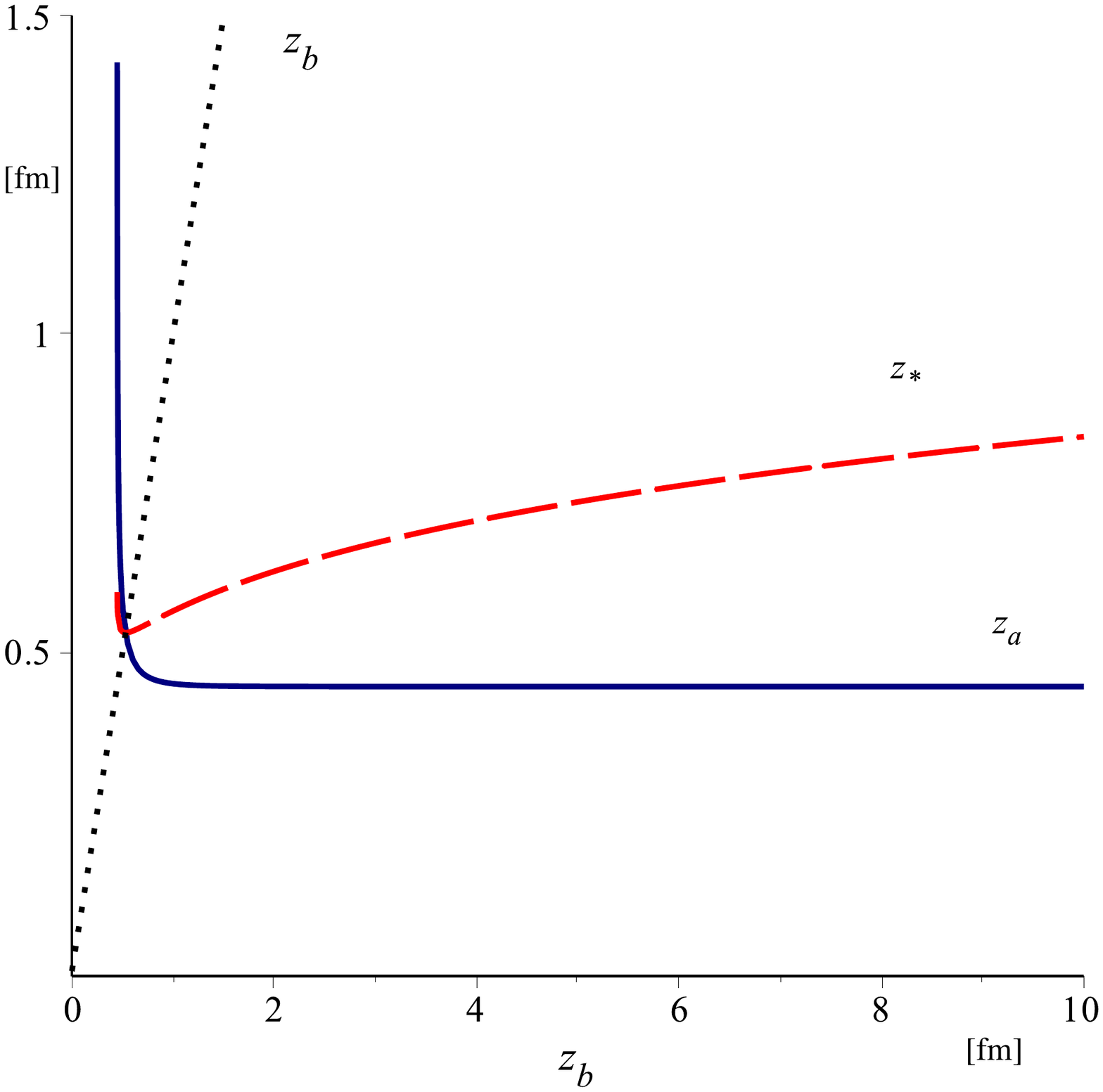}
 \caption{\label{fig:6}\textbf{The solution to the system of equations (\ref{d4d6zz}) for the given $z_{b}$.}}}
\end{figure}

Now we turn to the discussion of a wall-on-wall collision defined by the metric
\bea
ds^{2} = \frac{L^{2}}{z^{1/3}}\left[\frac{ \phi_{1}(z)}{z^{2}}\delta(u)du^{2} + \frac{ \phi_{2}(z)}{z^{2}}\delta(v)dv^{2}- \frac{dudv}{z^{2}} + \frac{dx^{2}_{2}}{z^{2}} + \frac{dz^{2}}{z^{2}} + \frac{\left(dy^{2}_{1} + dy^{2}_{2}\right)}{z^{4/3}}\right].\nn\\
\eea
The conditions for the trapped surface formation are
\bea\label{d4d6zz}
\frac{8\pi G_{5}Ez^{4}_{a}\left(1- \frac{z^{5}_{b}}{z^{5}_{*}}\right)}{\tilde{R}^{6}\left(\frac{z^{5}_{b}}{z^{5}_{*}} - \frac{z^{5}_{a}}{z^{5}_{*}} \right)} = - 1, \quad \frac{8\pi G_{5}Ez^{4}_{b}\left(1- \frac{z^{5}_{a}}{z^{5}_{*}}\right)}{\tilde{R}^{6}\left(\frac{z^{5}_{b}}{z^{5}_{*}} - \frac{z^{5}_{a}}{z^{5}_{*}} \right)} = 1,
\eea
which yield the expressions for the boundary and collision points
\be
z_{a} =\frac{z_{b}}{ \left(-1 +z^{4}_{b}C\right)^{1/4}}, \quad z_{*} = z^{4/5}_{a}z^{4/5}_{b}\left(\frac{z_{a}+z_{b}}{z^{4}_{a} + z^{4}_{b}}\right)^{1/5},
\ee
with
$C = 8\pi G_{5}E/\tilde{R}^{6}$. This solution is shown in Fig.~\ref{fig:6}.

The metric determinant for the $4d$ space 
\bea\label{8.2.5c}
ds^{2}_{Lif_{4}} =  \frac{\tilde{R}^{2}}{z^{1/3}}\left[ \frac{dx^{2}_{2}}{z^{2}} + \frac{\left(dy^{2}_{1} + dy^{2}_{2}\right)}{z^{4/3}} +  \frac{dz^{2}}{z^{2}} \right]
\eea
 is given by
\be \label{8.2.7}
\textrm{det}|g_{Lif_{4}}| = \frac{L^{8}}{z^{8}}.
\ee
Hence, for the relative entropy we have the following relation
\be \label{8.2.8d}
s = \frac{\tilde{R}^{4}}{6G_{5}}\left(\frac{1}{z^{3}_{a}} - \frac{1}{z^{3}_{b}}\right).
\ee
Taking $z_{b} \rightarrow \infty $ one comes to the following approximation 
\bea\label{8.2.9c}
s(C,z_{b}) &=& \left(\frac{C}{2}\right)^{3/4} - \left(\frac{1}{z_{b}}\right)^{3}  - \frac{3}{4}\left(\frac{2}{C}\right)^{1/4}\left(\frac{1}{z_{b}}\right)^{4} + \quad \ldots \quad. 
\eea
The maximum value of the trapped surface area at infinite $z_{b}$ is
\begin{equation}\label{8.9b}
 s|_{z_{b}\rightarrow \infty} =  \left(\frac{C}{2}\right)^{3/4} =
 \left(\frac{4\pi G_{5}}{\tilde{R}^{6}}\right)^{3/4} E^{3/4}.
\end{equation}

\subsection{From shock waves to $pp$-waves}

As it is known shock waves is a particular type of pp-waves, plane-fronted waves with parallel rays. 
These structures can be introduced in brane systems  to avoid divergences arising under the dimensional reduction.

If we consider the shock wave metric on  the $D4-D6$-brane intersection (\ref{8.2.4a}) taking some smooth function $f(t-x_{1})$ for the source of the shock wave instead of the $\delta$-function, one can define the function 
\begin{equation}\label{8.9pp}
K = 1 + \frac{\phi(z)}{z^{7/3}}f(t-x_{1}) ,
\end{equation}
where $\phi(z)$ is given by (\ref{8.2.8})-(\ref{8.2.8c}).  Now one  can turn the shock wave metric  into a more familiar form \cite{TO}
 \begin{eqnarray}\label{8.9d}
ds^{2} = \tilde{R}^{2}\left[ -K^{-1}dt^{2} +  K \left[dx_{1} + (K^{-1} - 1) dt \right]^{2} + \frac{dx^{2}_{2}}{z^{7/3}} + \frac{dz^{2}}{z^{7/3}} + \frac{\left(dy^{2}_{1} + dy^{2}_{2}\right)}{z^{4/3}}\right].
\end{eqnarray}

We note that $D4-D6$ can be obtained from the intersection of a magnetic $M5$-brane and the Kaluza-Klein monopole $\mathcal{KK}$ under the reduction over a relative transverse direction.
In \cite{Obers} gravitational waves propagating on the worldvolumes of intersecting branes have been used to construct Lifshitz spacetimes with hyperscaling violation using wave and transverse space reductions. Another way of generating Lifshitz solutions with hyperscaling violation in supergravities is  the dimensional reduction of null deformations of higher-dimensional branes defined on a product of manifolds, which includes $AdS$ spacetime  with wave structures  \cite{Narayan}.

\section{Conclusions}
\label{Sect:5}

In this paper, we have constructed shock wave geometries in Lifshitz-like ${\bf(p,q)}$ backgrounds (\ref{1.3}). 
We have considered Lifshitz-like metrics arising in four- and five-dimensional effective theories 
with a non-positive cosmological constant and gauge fields,  as well as in intersecting brane systems of supergravities  IIA and  IIB. 
We have found analytic solutions to the profile equations in the case of domain walls.
Following the holographic approach to estimate multiplicities produced in heavy-ion collisions,
we have considered the trapped surface formation for colliding domain walls.
We have obtained the conditions for the formation of trapped surfaces and calculated its areas.
In  $D = 5$ we have found  that  the results for the entropy in  the Lifshitz-like spacetimes type ${\bf(1,2)}$ with the exponent $\nu = 4$ 
and $\nu = -2$ for type ${\bf(2,1)}$, when the boundary points $z_{b}$ tend to its infinite values, match to that one calculated for the AdS-deformed background in \cite{APP}.  Note that the latter case corresponds to violating the null energy condition and thus this spacetime may not be physical.
Nonetheless, the Lifshitz geometry with $\nu = 4$ does not have the problem with the null energy condition in contrast 
to the background with the power-law factor $b(z)=\left(\frac{L_{eff}}{z}\right)^{1/2}$ from the work \cite{APP}, which provides the appropriate dependence of the entropy on the energy being supported by a phantom dilaton field.  The problem with the phantom nature of the dilaton field was one of the reason to consider the geometry with
$b(z)=\left(\frac{L_{eff}}{z}\right)^{1/2}$ only as an approximation to a true background in an intermediate zone
 \cite{Ageev:2014mma}.  From a general perspective we also can consider an anisotropic Lifshitz-like background as 
a suitable holographic background for a short time after collisions.

For wall-on-wall collisions defined in D3-D7 and D4-D6 brane systems, 
we have got the dependence of the entropy on energy as $E^{5/8}$ and $E^{3/4}$, respectively,  in the limit $z_{b} \rightarrow \infty$,
these predictions are not in  agreement with the experimental data.

There are several interesting  topics that are out of the scope of this paper and deserve further 
investigations. In the line of \cite{Ageev:2014mma}  we can find 
a relation between the entropy and the thermalization time assuming that the Lifshitz-like background
is just an approximated background in some intermediate zone. Then it would be interesting to see if these estimates  match to results obtained in
the Lifshitz-Vaidya model  \cite{KKVT}, where it was shown that the thermalization must take place at a finite velocities and with different values for different observables.

Another interesting extension of this work is to find solutions interpolating between AdS and Lifshitz-like backgrounds as it was done in   \cite{MT2},\cite{BGR}-\cite{BCG}. In particularly, it is important to study its finite-temperature flows and consider the evolution of the entropy during such isotropization model. Then it would be interesting to compare results with those expected in the Lifshitz hydrodynamics framework \cite{1304.7481}.

\acknowledgments
We would like to thank S. Cremonini for useful correspondences. 
This work was supported by the RFBR grant  14-01-00707. I.A. is also partially supported by
Program RAS P-19 "Fundamental problems of Nonlinear Dynamics".\\

\appendix

{\bf APPENDIX}

\section{Higher dimensional generalization} \label{App:A}

In this appendix, we will briefly discuss a shock wave construction in Lifshitz spacetimes of arbitrary dimensions.
As it was show in \cite{SSP3},  the Lifshitz-like metric  (\ref{2.1}) can be generalized to higher dimensions
\begin{equation}\label{A.01}
ds^{2} = L^{2}\left(r^{2\nu}\left(-dt^{2} + \sum^{p}_{i = 1}dx^{2}_{i}\right) + r^{2} \sum^{q}_{j = 1}dy^{2}_{j} + \frac{dr^{2}}{r^{2}}\right)
\end{equation}
with the anisotropic scaling 
\begin{equation}\label{A.02}
(t, x_{i}, y_{i}, r) \rightarrow \left(\lambda^{\nu}t, \lambda^{\nu}x_{i}, \lambda y_{i}, \frac{r}{\lambda}\right).
\end{equation}
In terms of the coordinate $\rho = r^{1/\nu}$ we rewrite (\ref{A.01}) as follows
\begin{equation}\label{A.03}
ds^{2} = L^{2}\left(\rho^{2}\left(-dt^{2} + \sum^{p}_{i = 1}dx^{2}_{i}\right) + \rho^{2/\nu} \sum^{q}_{j = 1}dy^{2}_{j} + \frac{d\rho^{2}}{\rho^{2}}\right).
\end{equation}
The corresponding Killing vectors are
\begin{eqnarray}\label{A.04}
\xi = -\frac{\partial}{\partial t}, \zeta_{1} =  \frac{\partial}{\partial x_{i}}, \zeta_{2} = \frac{\partial}{\partial y_{j}}, 
\eta = -t\frac{\partial}{\partial x_{i}} - x_{i} \frac{\partial}{\partial t}, 
 \chi = - x_{i} \frac{\partial}{\partial x_{i}} - t\frac{\partial}{\partial t} - \frac{y_{j}}{\nu}\frac{\partial}{\partial y_{j}} + \rho\frac{\partial}{d \rho}. \nonumber \\
\end{eqnarray}
The background (\ref{A.01}) in the holographic coordinate $z = 1/r$ takes the form
\begin{equation}\label{A.05}
ds^{2} = L^{2}\left(z^{-2}\left(-dudv + \sum^{p-1}_{i = 1}dx^{2}_{i}\right) + z^{-2/\nu} \sum^{q}_{j =1}dy^{2}_{j} + \frac{dz^{2}}{z^{2}}\right),
\end{equation}
where we introduce the null combinations $u = t - x_{p}$ and $v = t + x_{p}$.

Now we can write down the metric ansatz for the shock wave in the geometry (\ref{A.05})
\begin{equation}\label{A.06}
ds^{2} = L^{2}\left(z^{-2}\left( \phi(\vec{x}_{p-1}, \vec{y}_{q},z) du^{2} - dudv + \sum^{p-1}_{i = 1}dx^{2}_{i}\right) + z^{-2/\nu} \sum^{q}_{j = 1}dy^{2}_{j} + \frac{dz^{2}}{z^{2}}\right)
\end{equation}
with the profile function obeying the equation
\begin{equation}\label{A.07}
\left[\Box_{Lif_{(p+q)}} - \frac{1}{L^{2}}\left(p + \frac{q}{\nu}\right)\right]\frac{\phi(x_{i},y_{j},z)}{z} = -2z  J_{uu},
\end{equation}
where $i = 1, \ldots p-1, j  =1, \ldots q$,  $J_{uu}$  is the density related with stress-energy tensor, 
 and $\Box_{Lif_{(p+q)}}$ is the Laplace-Beltrami operator defined on the $(p+q)$-dimensional Lifshitz-like spacetime
\begin{equation}\label{A.08}
ds^{2}_{Lif_{(p+q)}} = \frac{1}{z^{2}}\sum^{p - 1}_{i =1}dx^{2}_{i} + \frac{1}{z^{2/\nu}}\sum^{q}_{j =1}dy^{2}_{j} + \frac{dz^{2}}{z^{2}}.
\end{equation}
The corresponding equation for the domain wall profile is reduced to the  form
\begin{equation}\label{A.09}
\frac{\partial^{2}\phi(z)}{\partial z^{2}} - \left(p + \frac{q}{\nu}\right)\frac{1}{z}\frac{\partial \phi(z)}{\partial z} = -16 \pi G_{D}Ez^{p + \frac{q}{\nu}}_{*}\delta(z - z_{*}).
\end{equation}
The solution to eq. (\ref{A.09}) can be represented in the form
\begin{equation}\label{A.010}
\phi = \phi_{a}\Theta(z_{*} - z) + \phi_{b}\Theta(z - z_{*}),
\end{equation}
with the profile functions given by
\begin{eqnarray}\label{A.011}
\phi_{a}(z) = C_{0}z_{a}z_{b}\left( \frac{z^{p+1 + \frac{q}{\nu}}_{*}}{z^{p+1 + \frac{q}{\nu}}_{b}} - 1\right)\left( \frac{z^{p+1 + \frac{q}{\nu}}}{z^{p+1 + \frac{q}{\nu}}_{a}}- 1\right),  \\
\phi_{b}(z) = C_{0}z_{a}z_{b}\left( \frac{z^{p+1 + \frac{q}{\nu}}_{*}}{z^{p+1 + \frac{q}{\nu}}_{a}} - 1\right)\left( \frac{z^{p+1 + \frac{q}{\nu}}}{z^{p+1 + \frac{q}{\nu}}_{b}}- 1\right),\\
C_{0} = -\frac{16\pi G_{D}E\nu z^{p+\frac{q}{\nu}}_{a}z^{p+\frac{q}{\nu}}_{b}}{(p\nu + q + \nu)L^{p+\frac{q}{\nu} +2}(z^{p+\frac{q}{\nu} +1}_{b} - z^{p+\frac{q}{\nu} +1}_{a})}.
\end{eqnarray}

The trapped surface is formed if the following conditions are satisfied
\begin{eqnarray}\label{A.012}
\frac{8\pi G_{(D)}E z^{p + \frac{q}{\nu}}_{a}\left(1 - z^{p + \frac{q}{\nu} +1}_{b}/z^{p + \frac{1}{\nu} +1}_{*}\right)}{L^{p+\frac{q}{\nu} +2 }\left(z^{p+ \frac{q}{\nu} + 1}_{b}/z^{p + \frac{q}{\nu} +1}_{*} - z^{p +\frac{q}{\nu} +1}_{a}/z^{p+ \frac{q}{\nu} +1}_{*}\right)} & = & -1, \nonumber\\  \label{A.3.5c}
\frac{8\pi G_{(D)}E z^{p + \frac{q}{\nu}}_{b}\left(1 - z^{p + \frac{q}{\nu} +1}_{b}/z^{p + \frac{q}{\nu} +1}_{*}\right)}{L^{p + \frac{q}{\nu} +2}\left(z^{p + \frac{q}{\nu} +1}_{b}/z^{p + \frac{q}{\nu} +1}_{*} - z^{p +\frac{q}{\nu} +1}_{a}/z^{p +\frac{q}{\nu} +1}_{*}\right)} &=&1.
\end{eqnarray}
The relations between the collision and boundary points read
\begin{equation}\label{A.013}
z_{a} = \left(\frac{z^{p+\frac{q}{\nu}}_{b}}{-1 + z^{p+\frac{q}{\nu}}_{b}C}\right)^{\frac{\nu}{\nu p +q}}, \quad z_{*} = \left(\frac{z^{p+\frac{q}{\nu}}_{a}z^{p+\frac{q}{\nu}}_{b}}{z^{p+\frac{q}{\nu}}_{a} + z^{p+\frac{q}{\nu}}_{b}}\right)^{\frac{\nu}{p\nu + \nu +q}}, 
\end{equation}
where $C = \displaystyle{\frac{8\pi G_{(D)}E}{L^{p + \frac{q}{\nu}+ 2}}}$.\\

The metric determinant for the $(p+q)$-dimensional Lifshitz space (\ref{A.08}) is
\begin{equation}\label{A.014}
\det{|g_{Lif_{(p+q)}}|} = z^{-(p + \frac{q}{\nu})}. 
\end{equation}

The relative entropy is given by
\begin{equation}\label{A.015}
s = \frac{1}{2G_{5}}\frac{\nu}{ \nu p + q - \nu}\left(\frac{z_{a}}{z^{p+\frac{q}{\nu}}}_{a} - \frac{z_{b}}{z^{p+\frac{q}{\nu}}_{b}} \right).
\end{equation}

\section{The shock wave and gauge fields}
In this section, we check that the gauge fields don't  contribute  to the profile's equation of the  shock wave. For simplicity, we show it for the $3+1$-dimensional model (\ref{2.1a}) with (\ref{2.0a}).
Let us write down the shock wave metric and the form fields in terms of light-cone coordinates (\ref{3.2})
\begin{equation}\label{D.1}
ds^{2} = \frac{\phi(y,z)\delta(u)}{z^{2}}du^{2} - \frac{1}{z^{2}} dudv + \frac{1}{z^{2/\nu}}\left(dy^{2}_{1} +dy^{2}_{2}\right) +  \frac{dz^{2}}{z^{2}},
\end{equation}
\begin{equation}\label{D.2}
H_{(3)} = - \sqrt{\frac{\nu -1}{\nu}}z^{-3}  du \wedge dv \wedge z, \quad B_{(2)} = \sqrt{\frac{\nu-1}{4\nu}}z^{-2} du\wedge dv.
\end{equation}
The Einstein equations corresponding to the geometry (\ref{2.1}) can be represented as
\begin{eqnarray}\label{D.3}
R_{MN} - \frac{1}{2}g_{MN}R + g_{MN}\Lambda = \frac{1}{4}H_{MM_{1}M_{2}}H_{N}^{M_{1}M_{2}} + m^{2}_{0}B_{MM_{1}}B_{N}^{M_{1}} -  \nn\\
\frac{1}{2}g_{MN}\left[\frac{H^{2}_{3}}{12} +\frac{m^{2}_{0}}{2}B^{2}_{2}\right].
\end{eqnarray}
The first component of eqs.(\ref{D.3}) is 
\begin{eqnarray}\label{D.4}
R_{uu} - \frac{1}{2}g_{uu}R = \frac{1}{4}H_{uvz}H_{u}^{\; {} v z} + m^{2}_{0}B_{u v}B_{u}^{\; {} v} - \frac{1}{2}g_{uu}\left[\frac{2\Lambda + H^{2}_{(3)}}{12} +\frac{m^{2}_{0}}{2}B^{2}_{2}\right],
\end{eqnarray}
where 
\begin{equation}\label{D.5}
H_{u v z}H_{u}^{ \; {} v z} =  H_{u v z}H_{uv z} g^{vv}g^{zz}, \quad  B_{u v}B_{u}^{\; {}v} = B_{u v} B_{uv}g^{vv}.
\end{equation}
The non-diagonal component of the Einstein equations reads
\begin{eqnarray}\label{D.6}
R_{uv} - \frac{1}{2}g_{uv}R = \frac{1}{4}H_{uv z}H_{v}^{\; {} v z} + m^{2}_{0}B_{uv}B_{v}^{\; {} v} - \frac{1}{2}g_{uv}\left[\frac{2\Lambda + H^{2}_{3}}{12} +\frac{m^{2}_{0}}{2}B^{2}_{2}\right],
\end{eqnarray}
where 
\begin{equation}\label{D.7}
H_{uv z}H_{v}^{\; {} v z} =  H_{uv z}H_{v u z} g^{ u v}g^{zz}, \quad  B_{uv}B_{v}^{\; {} v} = B_{uv} B_{vu}g^{uv}.
\end{equation}

Inserting the metric components  $g_{uu}$, $g^{vv}$  in (\ref{D.5}) and $g_{uv}$, $g^{uv}$  in (\ref{D.6}) one comes to
\begin{eqnarray}\nn
R_{uu} - \frac{1}{2}g_{uu}R &=&  -  \phi(y ,z)\delta(u)\left[4z^{2}\left(\frac{H_{uvz}H_{uv z}g^{zz}}{4} + m^{2}_{0}B_{uv}B_{uv}\right) + \frac{\left(2\Lambda + \frac{H^{2}_{3}}{12} + \frac{m^{2}_{0}}{2}B^{2}_{2}\right)}{2z^{2}}\right],\\ \label{D.5b}\\
R_{uv} - \frac{1}{2}g_{uv}R &=&2z^{2}\left(\frac{H_{uvz}H_{uv z}g^{zz}}{4} + m^{2}_{0}B_{uv}B_{uv}\right)+ \frac{1}{4z^{2}}\left(2\Lambda + \frac{H^{2}_{3}}{12} + \frac{m^{2}_{0}}{2}B^{2}_{2}\right).\nn\\
\label{D.5c}
\end{eqnarray}
Substituting the left-hand side of (\ref{D.5c}) into the right-hand side of (\ref{D.5b}) we obtain 
\begin{equation}\label{D.8}
R_{uu} - \frac{1}{2}g_{uu}R =  - 2\phi(y,z)\delta(u)\left[ R_{uv} - \frac{1}{2}g_{uv}R \right].
\end{equation}
Taking into account the expression for the scalar curvature (\ref{3.7a}) one has
\begin{eqnarray}\label{D.8b}
 \frac{1}{2}g_{uu}R =  - \phi(y,z)\delta(u) \frac{(3\nu^{2} + 2\nu + 1)}{z^{2} \nu^{2}} \quad \text{and}\quad \nn\\
 g_{uv}R\phi(y,z)\delta(u)  =  \phi(y,z)\delta(u)\frac{(3\nu^{2} + 2\nu + 1)}{z^{2} \nu^{2}}.
\end{eqnarray}
Thus, owing to (\ref{D.8b}) eq. (\ref{D.8}) can be represented in the following form
\begin{equation}\label{D.9}
R_{uu} +2 \phi(y,z)\delta(u) R_{uv} = 0.
\end{equation}
Substituting relations for $R_{uu}$ and $R_{uv}$ in eq. (\ref{D.9}) we have
\begin{eqnarray}\label{D.9a}
 - \frac{1}{2}\frac{\delta(u)}{z\nu}\Bigl[z\nu\frac{\partial^{2} \phi(y,z)}{\partial z^{2}} +  z^{-1+ \frac{2}{\nu}}\nu\frac{\partial^{2}\phi(y,z)}{\partial y^{2}} - \frac{\partial \phi(y,z)}{\partial z}\nu  -  
 \frac{\partial \phi(y,z)}{\partial z} + \nonumber \\ \frac{4\phi(y,z) \nu}{z} + \frac{2\phi(y,z)}{z}  \Bigr]   + \phi(y,z)\delta(u)\frac{2\nu + 1}{\nu z^{2}} = 0.
\end{eqnarray}
In the left-hand side of eq. (\ref{D.9a}) we can derive the relation for the operator $\Box_{Lif_{2}}$ (\ref{3.6a}) and represent the equation in the form
\begin{equation}\label{D.9b}
\delta(u)\left[ \Box_{Lif_{2}} -  \left(1 + \frac{1}{\nu}\right)\right]\frac{ \phi(y,z)}{z}  = 0.
\end{equation}
Thus, we can conclude that $H_{(3)}$ and $B_{(2)}$ do not give a contribution to the equations for the shock wave profile.

\section{The profile equations for shock waves in $D = 5$}

In this section, we derive  equations for the shock wave profiles in D=5 $Lif_{(p,q)}$ spacetimes.

\subsection{ $Lif_{(1,2)}$-type} \label{App:B1}
Here we consider the shock-wave in the Lifshitz-like geometry ${\bf(1,2)}$  with the metric (\ref{4.1.4a}). 
To show that the profile equation is given by (\ref{4.1.13}), we write down corresponding non-zero components of the Ricci tensor 
\begin{eqnarray}\label{B.1}
R_{uu} = - \frac{1}{2}\frac{\delta(u)}{z\nu}\Bigl[z\nu\frac{\partial^{2} \phi(y_{1},y_{2},z)}{\partial z^{2}} +  z^{-1+ \frac{2}{\nu}}\nu\frac{\partial^{2}\phi(y_{1},y_{2},z)}{\partial y_{1}^{2}} + z^{-1+ \frac{2}{\nu}}\nu \frac{\partial^{2} \phi(y_{1},y_{2},z)}{\partial y_{2}^{2}}  \nonumber \\
- \frac{\partial \phi(y_{1},y_{2},z)}{\partial z}\nu  -  2\frac{\partial \phi(y_{1},y_{2},z)}{\partial z} + \frac{4\phi(y_{1},y_{2},z) \nu}{z} + \frac{4\phi(y_{1},y_{2},z)}{z}  \Bigr].
\end{eqnarray}
\begin{eqnarray}\label{B.2}
R_{uv} = \frac{(\nu+ 1)}{\nu z^{2}}, \quad  R_{y_{1}y_{1}} = R_{y_{2}y_{2}} = - \frac{2z^{-\frac{2}{\nu}}(\nu + 1)}{\nu^{2}}, \quad R_{zz} = - \frac{2(\nu^{2} + 1)}{\nu^{2}z^{2}}.
\end{eqnarray}
The scalar curvature reads
\begin{eqnarray}\label{B.3}
R = - \frac{2(3\nu^{2} + 4\nu +3)}{L^{2} \nu^{2}}.
\end{eqnarray}
The equation for the profile the $(u,v)$ component of the Einstein equations reads
\begin{equation}\label{B.4}
R_{uv} - \frac{1}{2}g_{uv}R + g_{uv}\Lambda = T_{uv},
\end{equation}
where  $T_{uv} = 0$, since the only non-zero component of the stress-energy tensor is $T_{uu}$.
Substituting $R_{uv}$, $g_{uv}$ into (\ref{B.4})  we find the relation for $\Lambda$
\begin{equation}\label{B.6}
\Lambda =  - \frac{1}{L^{2}}\left(1 + \frac{2}{\nu} + \frac{3}{\nu^{2}}\right).
\end{equation}
Inserting $R_{uu}$ from (\ref{B.1}), $g_{uu}$ from (\ref{4.1.4a}), (\ref{B.3}) and (\ref{B.6}) into the $(u,u)$-component, we find
\begin{eqnarray}\label{B.7}
 - \frac{1}{2}\frac{\delta(u)}{z\nu}\Bigl[z\nu\frac{\partial^{2} \phi(y_{1},y_{2},z)}{\partial z^{2}} +  z^{-1+ \frac{2}{\nu}}\nu\frac{\partial^{2}\phi(y_{1},y_{2},z)}{\partial y_{1}^{2}} + z^{-1+ \frac{2}{\nu}}\nu \frac{\partial^{2} \phi(y_{1},y_{2},z)}{\partial y_{2}^{2}}  \nonumber \\
- \frac{\partial \phi(y_{1},y_{2},z)}{\partial z}\nu  -  2\frac{\partial \phi(y_{1},y_{2},z)}{\partial z} + \frac{4\phi(y_{1},y_{2},z) \nu}{z} + \frac{4\phi(y_{1},y_{2},z)}{z}  \Bigr] + \nn\\
\delta(u) \frac{\phi(y_{1},y_{2},z)}{z^{2}}\frac{(3\nu^{2} + 4\nu +3)}{\nu^{2}}  - \delta(u)\frac{\phi(y_{1},y_{2},z)}{z^{2}}\left(1 + \frac{2}{\nu} + \frac{3}{\nu^{2}}\right) = T_{uu}.
\end{eqnarray}
Simplifying (\ref{B.7}) and taking into account the expression for the Laplace-Beltrami operator (\ref{4.1.6b}) one comes to eq. (\ref{4.1.13}).

\subsection{$Lif_{(2,1)}$-type} \label{App:B2} 
Now we turn to the shock wave metric (\ref{4.4.4}) with the following non-zero components of the Ricci tensor 
\begin{eqnarray}\label{B.8}
R_{uu} = - \frac{1}{2}\frac{\delta(u)}{\nu}\Bigl[\nu\frac{\partial^{2} \phi(x_{2},y_{2},z)}{\partial z^{2}} +  \nu\frac{\partial^{2}\phi(x_{2},y_{2},z)}{\partial x_{2}^{2}} + z^{-2+ \frac{2}{\nu}}\nu \frac{\partial^{2} \phi(x_{2},y_{2},z)}{\partial y_{2}^{2}}  \nonumber \\
- \frac{2\nu}{z}\frac{\partial \phi(x_{2},y_{2},z)}{\partial z}  -  \frac{1}{z}\frac{\partial \phi(x_{2},y_{2},z)}{\partial z} + \frac{6\phi(x_{2},y_{2},z) \nu}{z^{2}} + \frac{2\phi(x_{2},y_{2},z)}{z^{2}} \Bigr],
\end{eqnarray}
\begin{eqnarray}\label{B.9}
R_{uv} =\frac{3\nu+1}{2 \nu z^{2}}, \quad  R_{x_{2}x_{2}} = - \frac{3\nu+1}{\nu z^{2}},   \quad R_{y_{2}y_{2}} =- z^{-2/\nu}\frac{3\nu+1}{\nu^{2}}, \quad R_{zz} =- \frac{3\nu^2+1}{\nu^{2}z^{2}}.
\end{eqnarray}
The scalar curvature corresponding to  (\ref{4.4.4}) reads
\begin{eqnarray}\label{B.10}
R = - \frac{2(6\nu^2+3\nu+1)}{L^{2}\nu^2}.
\end{eqnarray}
A non-diagonal component of Einstein equations reads
\begin{equation}\label{B.11}
R_{uv} - \frac{1}{2}g_{uv}R + g_{uv}\Lambda = T_{uv}.
\end{equation}

Substituting $R_{uv}$, $R$  and $g_{uv}$ into the non-diagonal component of the Einsten equations, one obtains
\begin{equation}\label{B.12}
\frac{3\nu+1}{\nu z^{2}} - \frac{1}{4z^{2}}\frac{2(6\nu^{2} + 3\nu + 1)}{\nu^{2}} - \frac{\Lambda}{2z^{2}} = 0,
\end{equation}
which yields the relation for $\Lambda$
\begin{equation}\label{B.13}
\Lambda = - \frac{1}{L^{2}}\left( 3 + \frac{2}{\nu} + \frac{1}{\nu^{2}}\right).
\end{equation}
The $uu$-component of Einstein equations is given
\begin{eqnarray}\label{B.14}
- \frac{1}{2}\frac{\delta(u)}{\nu}\Bigl[\nu\frac{\partial^{2} \phi(x_{2},y_{2},z)}{\partial z^{2}} +  \nu\frac{\partial^{2}\phi(x_{2},y_{2},z)}{\partial x_{2}^{2}} + z^{-2+ \frac{2}{\nu}}\nu \frac{\partial^{2} \phi(x_{2},y_{2},z)}{\partial y_{2}^{2}}  \nonumber \\
- \frac{2\nu}{z}\frac{\partial \phi(x_{2},y_{2},z)}{\partial z}  -  \frac{1}{z}\frac{\partial \phi(x_{2},y_{2},z)}{\partial z} + \frac{6\phi(x_{2},y_{2},z) \nu}{z^{2}} + \frac{2\phi(x_{2},y_{2},z)}{z^{2}} \Bigr] + \nn\\
\frac{\delta(u) \phi(x_{2},y_{2},z)}{z^{2}} \frac{(6\nu^2+3\nu+1)}{\nu^{2}} - \delta(u) \frac{ \phi(x_{2},y_{2},z)}{z^{2}} \left(3 + \frac{2}{\nu} + \frac{1}{\nu^{2}}\right) = T_{uu},
\end{eqnarray}
which yields eq. (\ref{4.4.10}).

\section{The  curvature invariants}
Here we show that the curvature invariants for the Lifshitz-like background (\ref{1.3}) have  finite values.
For simplicity, we consider the 5-dimensional Lifshitz-like metric type ${\bf(1,2)}$ 

\begin{equation}\label{A.1}
ds^{2} = \frac{\left(-dt^{2} + dx^{2}\right)}{z^{2}} + \frac{\left(dy^{2}_{1} + dy^{2}_{2}\right)}{z^{2/\nu}} +  \frac{dz^{2}}{z^{2}}.
\end{equation}

Nonzero components of the Riemann tensor corresponding to (\ref{A.1}) are 
\begin{eqnarray}\label{A.3}
R_{txtx} =  \frac{1}{z^{4}},\quad R_{ty_{1}ty_{1}} = \frac{z^{-2/\nu}}{\nu z^{2}},\quad R_{ty_{2}ty_{2}} =  \frac{z^{-2/\nu}}{\nu z^{2}}, \quad R_{tztz} =  \frac{1}{z^{4}}, \\
 R_{xy_{1}xy_{1}} = - \frac{z^{-2/\nu}}{\nu z^{2}},  \quad R_{xy_{2}xy_{2}} =  -\frac{z^{-2/\nu}}{\nu z^{2}}, \quad R_{xzxz} = -\frac{1}{z^{4}}, \quad R_{y_{1}y_{2}y_{1}y_{2}} = - \frac{z^{-4/\nu}}{\nu^{2}}, \nn\\
 R_{y_{1}zy_{1}z}  =   - \frac{z^{-2/\nu}}{\nu^{2}z^{2}}, \quad R_{y_{2}zy_{2}z} =  - \frac{z^{-2/\nu}}{\nu^{2} z^{2}}. \nn
\end{eqnarray}
Raising indices in (\ref{A.3}) we get
\begin{eqnarray}\label{A.4}
R^{txtx} = z^{4}, \quad R^{ty_{1}ty_{1}} = \frac{1}{\nu}z^{2+2/\nu}, \quad  R^{ty_{2}ty_{2}} = \frac{1}{\nu}z^{2+2/\nu}, \quad R^{tztz} =  z^{4}, \\
 R^{xy_{1}xy_{1}} = -\frac{1}{\nu}z^{2+2/\nu}, \quad  R^{xy_{2}xy_{2}} = -\frac{1}{\nu}z^{2+2/\nu}, \quad R^{xzxz} = - z^{4}, \nn\\
  R^{y_{1}y_{2}y_{1}y_{2}} = -\frac{1}{\nu^{2}}z^{4/\nu},
 R^{y_{1}zy_{1}z}  =   - \frac{z^{2 +2/\nu}}{\nu^{2}}, \quad R_{y_{2}zy_{2}z} =  - \frac{z^{2+2/\nu}}{\nu^{2}}.\nn
\end{eqnarray}

The scalar curvature for (\ref{A.1}) is
\begin{equation}\label{A.4a}
R = -  \frac{2\left(3\nu^{2} + 4\nu  + 3\right)}{\nu^{2}}.
\end{equation}

 The Kretschmann scalar  can be represented as
\begin{eqnarray}\label{A.5}
K  = R_{abcd}R^{abcd} = R_{txtx}R^{txtx} + R_{ty_{1}ty_{1}}R^{ty_{1}ty_{1}} +  R_{ty_{2}ty_{2}}R^{ty_{2}ty_{2}} + R_{tztz}R^{tztz}+ \nonumber \\
R_{xy_{1}xy_{1}}R^{xy_{1}xy_{1}}+R_{xy_{2}xy_{2}} R^{xy_{2}xy_{2}} +  R_{xzxz} R^{xzxz} + R_{y_{1}y_{2}y_{1}y_{2}}R^{y_{1}y_{2}y_{1}y_{2}} + \nonumber \\
R_{y_{1}zy_{1}z} R^{y_{1}zy_{1}z} +R_{y_{2}zy_{2}z}R^{y_{2}zy_{2}z}.
\end{eqnarray}
Substituting the Riemann tensor components from (\ref{A.3}) and (\ref{A.4}) into (\ref{A.5}) we have
\begin{eqnarray}
K  =  \frac{1}{z^{4}} z^{4} +  \frac{z^{-2/\nu}}{\nu z^{2}} \frac{1}{\nu}z^{2+2/\nu} + \frac{z^{-2/\nu}}{\nu z^{2}}\frac{1}{\nu}z^{2+2/\nu} + \frac{1}{z^{4}} z^{4} + \nonumber \\
\left( - \frac{z^{-2/\nu}}{\nu z^{2}}\right)\left( -\frac{1}{\nu}z^{2+2/\nu}\right) + \left( -\frac{z^{-2/\nu}}{\nu z^{2}}\right)\left( -\frac{1}{\nu}z^{2+2/\nu}\right)+\left( -\frac{1}{z^{4}}\right)\left( - z^{4}\right) + \nonumber\\
 \left( - \frac{z^{-4/\nu}}{\nu^{2}}\right)\left(-\frac{1}{\nu^{2}}z^{4/\nu}\right) +
 \left( - \frac{z^{-2/\nu}}{\nu^{2}z^{2}}\right)\left( - \frac{z^{2 +2/\nu}}{\nu^{2}}\right) + \left( - \frac{z^{-2/\nu}}{\nu^{2} z^{2}}\right)\left( - \frac{z^{2+2/\nu}}{\nu^{2}}\right) = \nonumber\\
\frac{3+ 4\nu^{2} + 3\nu^{4}}{\nu^{4}}.
\end{eqnarray}

Now let us show that   $\nabla_P R_{mnkl}\nabla^P R^{mnkl}$ is also finite. It can be written down in the detailed form
\begin{eqnarray}\label{A.6}
 \nabla_P R_{mnkl}\nabla^P R^{mnkl} = \nabla_z R_{txtx}\nabla^z R^{txtx} +  \nabla_z R_{ty_{1}ty_{1}}\nabla^z R^{ty_{1}ty_{1}} + \nabla_{z}R_{ty_{2}ty_{2}}\nabla^{z}R^{ty_{2}ty_{2}} + \nonumber \\ \nabla_{z}R_{tztz}\nabla^{z}R^{tztz} +  \nabla_{z}R_{xy_{1}xy_{1}}\nabla_{z}R^{xy_{1}xy_{1}} + \nabla_{z} R_{xy_{2}xy_{2}}\nabla^{z} R^{xy_{2}xy_{2}} + \nabla_{z}R_{xzxz}\nabla^{z} R^{xzxz} + \nonumber \\
\nabla_{z}R_{y_{1}y_{2}y_{1}y_{2}}\nabla^{z}R^{y_{1}y_{2}y_{1}y_{2}} + \nabla_{z}R_{y_{1}zy_{1}z} \nabla^{z}R^{y_{1}zy_{1}z} +  \nabla_{z}R_{y_{2}zy_{2}z}\nabla^{z}R^{y_{2}zy_{2}z}.\nn\\
\end{eqnarray}
For the first term we have
\begin{eqnarray}\label{A.5a}
\nabla_z R_{txtx} = R_{txtx, z} - R_{txtx}\Gamma^{t}_{tz} - R_{txtx}\Gamma^{x}_{xz} - R_{txtx}\Gamma^{t}_{tz} - R_{txtx}\Gamma^{x}_{xz} = \nonumber\\
-\frac{4}{z^{5}} +  \frac{1}{z^{5}} + \frac{1}{z^{5}} + \frac{1}{z^{5}} + \frac{1}{z^{5}} = 0.
\end{eqnarray}
The other covariant derivatives are also equal to zero
\begin{eqnarray}\label{A.6b}
 \nabla_z R_{ty_{1}ty_{1}}= 0, \quad  \nabla_{z}R_{ty_{2}ty_{2}} = 0, \quad \nabla_{z}R_{tztz} = 0, \quad  \nabla_{z}R_{xy_{1}xy_{1}} = 0, \nonumber\\
  \nabla_{z} R_{xy_{2}xy_{2}} = 0, \quad  \nabla_{z}R_{xzxz} = 0, \quad \nabla_{z}R_{y_{1}y_{2}y_{1}y_{2}} = 0, \nonumber\\
    \nabla_{z}R_{y_{1}zy_{1}z} = 0, \quad   \nabla_{z}R_{y_{2}zy_{2}z} =0. 
\end{eqnarray}

Hence, taking  (\ref{A.5a}) and (\ref{A.6b}) into account, we obtain that
\begin{equation}\label{A.6a}
 \nabla_P R_{mnkl}\nabla^P R^{mnkl} = 0.
\end{equation}

\end{document}